\documentclass{article}
\usepackage{amsmath}
\usepackage{algorithmic}
\usepackage{graphicx}
\usepackage{textcomp}
\usepackage{xcolor}
\usepackage{fancyhdr}
\usepackage{float}
\usepackage[caption = false]{subfig}
\usepackage[hyphens]{url}
\usepackage[left=2cm,right=2cm]{geometry}
\renewcommand{\b}{\textcolor{black}}
\usepackage{authblk}

\usepackage{braket}
\usepackage{soul}

\title{ Benchmarking Quantum Computers and the Impact of Quantum Noise }
\author{
  Salonik Resch \;\;\;\;\;\;\;\;\;\;\;\;\;\;\;\;Ulya R. Karpuzcu \\
  \texttt{resc0059@umn.edu}\;\;\;\;\;\;\;\;\;\;\texttt{ukarpuzc@umn.edu}
}
\date{University of Minnesota\\May 2021}

\begin{document}

\maketitle

\begin{abstract}
    Benchmarking is how the performance of a computing system is determined.  Surprisingly, even for classical computers this is not a straightforward process. One must choose the appropriate benchmark and metrics to extract meaningful results. Different benchmarks test the system in different ways and each individual metric may or may not be of interest. Choosing the appropriate approach is tricky. The situation is even more open ended for quantum computers, where there is a wider range of hardware, fewer established guidelines, and additional complicating factors. Notably, quantum noise significantly impacts performance and is difficult to model accurately. Here, we discuss benchmarking of quantum computers from a computer architecture perspective and provide numerical simulations highlighting challenges which suggest caution.  
\end{abstract}

\section{Introduction}
There are many ways to measure the performance of a computer\footnote{For clarification, we note benchmarks are operations that the system is asked to perform (programs) and metrics are measurable characteristics of the system when performing the benchmark.}. Common ways have been measuring \textit{operations per second} (OPS) or \textit{floating-point operations per second} (FLOPS). These are intuitive and easy to understand, however, they are generally poor metrics. 
The problem is that hardware could be designed to have a very high OPS/FLOPS but could perform poorly on real world applications, which do not consist of monolithic blocks of arithmetic operations. A way to improve upon this is to measure the progress of a program rather than the number of operations it performs. For example, the HINT benchmark measures quality improvements per second (QUIPS), which measures the numerical accuracy improvement of the output in a given time \cite{gustafson1995hint}. While this can be insightful, again the main concern is that this does not accurately represent the real-world programs that will be run on the hardware. 

Generally, the best metric is the wall-time required to complete a program \cite{lilja2005measuring}, if the program is representative of real-world applications. This concept has led to the creation of benchmarks which are samples of larger, industrially useful applications. SpecCPU \cite{spradling2007spec} and Parsec \cite{bienia2008parsec} are popular suites in this vein. While this is a clear improvement, it is not without its issues. For one, the reduced size of the programs introduces estimation error on the performance. There are other, less obvious complications. For example, academic work commonly reports performance on Parsec for system evaluation. Now, the human made choices of which benchmarks to include in Parsec determine what the academic community considers to be important. This makes these choices critical, because if 
the selection 
%Parsec 
is not representative these results can be misleading. Even further, having established benchmarks would allow for a hardware designer to ``cheat'' by making a system particularly good on only the specific applications.

The takeaway is that benchmarking is possible and useful, yet is tricky and can be misleading. It is difficult to create useful benchmarks, and it may be impossible to create universal ones. This same construct applies to quantum computing, except it is much more intricate. There are a number of complicating factors:
\begin{list}{\labelitemi}{\leftmargin=1em}
%\begin{enumerate}
    \item[1)] Quantum hardware is more diverse than classical hardware;
    \item[2)] Quantum hardware is less developed, most systems have only a few qubits and cannot perform useful applications;
    \item[3)] Quantum algorithms are still being developed and it is unknown what applications will be the most useful;
    \item[4)] Quantum noise is not well understood and difficult to simulate, making characterization particularly challenging.
%\end{enumerate}
\end{list}
The dissimilarity of quantum hardware makes it hard to compare them to each other. This is not as much of an issue for classical hardware. While there has been a trend towards more diversified and specialized hardware in recent years, such as application specific integrated circuits (ASIC), there is a general framework and almost all hardware is silicon CMOS based. This makes metrics, benchmarks, and general intuition portable across different devices. Currently, there are many different hardware approaches competing in quantum computing. Each is based on a different physical system with entirely different dynamics. For example, quantum computing can be performed in superconducting circuits, ions isolated in a vacuum, or in atoms embedded in silicon. These systems look very different from each other and each have unique advantages and deficiencies. Is it fair to compare them directly?

As quantum computing is early on in its development, there are only small-to-medium sized quantum computers in existence. Most systems are not capable of performing useful programs. This makes it difficult to create benchmarks for these systems that are representative of future real-world applications. Scaling to larger sizes is particularly difficult for quantum computers, hence benchmarks that can be run on these smaller systems are less likely to accurately represent the performance of scaled-up versions. This is where one would normally turn to simulation. Unfortunately, as the states in quantum computers are highly complex, they are not able to be efficiently simulated by classical computers. Hence, benchmarks must be tied to a physical experiment.

On a more fundamental level, it is even unsure what quantum applications will be useful. As the field of quantum computing is largely unexplored, and not well understood, it is believed that many of its advantages and potential are currently unknown. Exploration of quantum potential is not well captured by benchmarking \cite{blume2019metrics}.

Quantum computing faces many hardware challenges. Information is easily lost due to quantum noise, which causes decoherence of quantum states. The physical devices need near absolute isolation from the environment, making the systems large and difficult to scale. Due to this fragility, benchmarking begins much lower in the system stack.  Benchmarks even for 1-bit operations have been developed \cite{knill2008randomized,rb2009,rb2005}. Even at this level, performance has been difficult to quantify. Accurately modeling quantum noise and determining the robustness of quantum operations has become the subject of much research \cite{wallman2016noise,erhard2019characterizing,knill2008randomized,proctor2017randomized}. Noise can affect quantum programs differently, depending on their length and structure. Hence, noise is a significant complicating factor.

Thus, quantum computing inherits all the benchmarking complexity of classical computing, but introduces many additional complications. This makes it quite unclear what the best way is to evaluate a quantum system. In fact, the authors of \cite{blume2019metrics} argue that it is too early to develop a standard approach. They warn that quantum research is currently exploratory in nature and that benchmarks are inappropriate for this kind of work. In fact, it could even be detrimental due to the possibility of misguiding research efforts.

In this paper we provide a quantitative comparison 
%and contrast 
for 
%Here, we discuss 
different quantum strategies from a benchmarking perspective, in the presence of quantum noise, to pinpoint pitfalls and fallacies. 
%and provide numerical results that suggest caution. 
We start with basics in Section \ref{sec:primer}. We discuss quantum noise and noise models in Section \ref{sec:noise}. A brief introduction to quantum metrics is supplied in Section \ref{sec:metrics}. Benchmarking for single- and two-qubit systems is covered in Section \ref{sec:qubit}, for near-term noisy computing systems in Section \ref{sec:computer}, and for fault-tolerant computing systems in Section \ref{sec:faulttolerant}. Sections \ref{sec:experiments} and \ref{sec:results} detail the simulation results. Finally, we conclude the paper in Section \ref{sec:conc}.
%and discuss the results in Section \ref{sec:results}. Finally, we set up some experiements in Section \ref{sec:experiments} and discuss the results in Section \ref{sec:results}.

\section{Quantum Primer}
\label{sec:primer}
In this section we introduce background and context for quantum computing. Necessarily brief, this clearly cannot do it justice. Quantum mechanics is highly complex and defies intuition. To quote Richard Feynman, ``If you think you understand quantum mechanics, you don't understand quantum mechanics.'' This seems even more applicable if one views quantum mechanics from the perspective of computer architecture \cite{hennessy2011computer}. But in an attempt to ``understand quantum mechanics'', we will attempt to cover key concepts. We recommend \cite{griffiths2003consistent} as an introduction to quantum mechanics and \cite{loceff2015course} as an introduction to quantum computing.

Quantum mechanics describes the nature of the physical world. High temperatures and large sizes causes quantum mechanical effects to become less noticeable, and classical physics acts as a good approximation. But when one creates a system that is very small or very cold, only quantum mechanics can accurately describe the system and how it evolves in time. Under these conditions, states are noticeably quantized (take on discrete values), such as the discrete possible energy levels of electrons around the nucleus of atoms. We can assign logical values to these distinct states, which are then called \emph{qudits}. Transitions between these states correspond to quantum logical operations. Qudits can have many possible values, for example there are many possible non-degenerate energy levels for an electron. However, it is often convenient to use only two of the possible states, such as the ground and first excited states, as these become analogous to classical bits and are less susceptible to noise \cite{nielsen2002quantum}. These two-level qudits are called \emph{qubits}. Qubits can be in both of their states simultaneously (superposition) and multiple qubits can have their states intertwined (entanglement). Hence, there is not only information in each qubit, but \emph{between} each qubit. As a direct consequence, quantum states can store an amount of information that is exponential in the number of qubits. This enables extreme compute capabilities if one is able to create a complex quantum state and reliably transform it in a meaningful way. Unfortunately, this is a difficult task. Pure \footnote{Pure states are quantum states which can be completely specified by state vectors.} quantum states are extremely fragile and need near perfect isolation from the environment to exist. At the same time, we need to be able to interact with the quantum state in order to transform it.

Large scale quantum computing, despite the fragility of quantum states, remains a possibility due to quantum error correction (QEC). By encoding quantum information for a single qubit using multiple qubits, the quantum state can be restored if only a subset of the qubits become corrupted. Encoded qubits are called \emph{logical} qubits, which are composed of multiple \emph{physical} qubits. QEC is a rich field \cite{gottesman2010introduction}, and there is much work devoted to studying how QEC works under different error models \cite{gutierrez2015comparison,gutierrez2013approximation,bravyi2018correctingcoherenterrors,greenbaum2017modelingcoherent,magesan2013modeling}. However, modern quantum computers do not yet have sufficient numbers of qubits or required qubit quality to practically implement QEC. Hence, modern quantum applications operate on physical qubits and try to make use of limited resources. Therefore, it is of interest not only how quantum error affects QEC, but also how it affects algorithms running without QEC.

\begin{figure}
%\vspace{.2cm}
    \centering
    \includegraphics[scale=0.23]{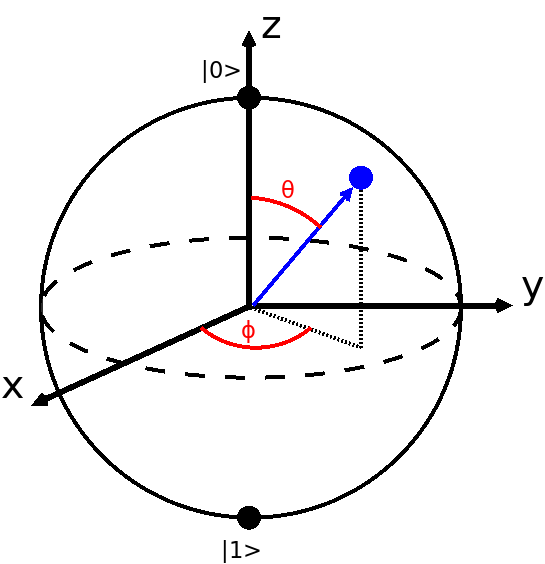}
 %   \vspace{.1cm}
    \caption{Bloch Sphere representation of a single qubit.}
    \label{fig:blochsphere}
%    \vspace{.4cm}
\end{figure}

%\subsection{Terminology}
 If a quantum state is completely isolated, it is called a \emph{pure state}. Quantum pure states can be represented by \emph{kets}, which are column vectors of complex numbers. The elements of the kets are called the \emph{amplitudes}. For example, a single qubit can be represented by a ket of length 2
 %\vspace{.1cm}
 \begin{equation}
\alpha \ket 0 + \beta \ket 1 = \left [ \begin{matrix} \alpha \\ \beta \end{matrix} \right ]
\end{equation}
where $\alpha$ is the amplitude associated with state $\ket 0$ and $\beta$ with state $\ket 1$. The amplitudes determine probabilities when performing measurements, $|\alpha|^2$ is the probability of measuring this single qubit in the state $\ket 0$ \b{and $|\beta|^2$ is the probability of measuring the qubit in the $\ket 1$ state.} The probabilities must sum to 1 for pure states. A single qubit can be visualized as a vector from the origin to a point on the Bloch Sphere, shown in Figure \ref{fig:blochsphere}, where a pair of angles, $\theta$ and $\phi$, can be used to specify the state of the qubit, related to the amplitudes by the equations
%\vspace{.5cm}
\begin{equation}\alpha = cos(\theta /2)\end{equation}
\begin{equation} \beta = e^{i\phi}sin(\theta/2)\end{equation}

Operations on qubits are called \emph{gates}, which are represented by matrices. Common gates are the Pauli \textbf{I} (identity), \textbf{X} (NOT), \textbf{Z} (phase-flip), and \textbf{Y} (NOT and phase flip) gates, \b{shown in Equation \ref{eq:matrices}}. 
\begin{equation} \b{\textbf{I} = \left [ \begin{matrix} 1 & 0 \\ 0 & 1 \end{matrix} \right ]}\qquad \textbf{X} = \left [ \begin{matrix} 0 & 1 \\ 1 & 0 \end{matrix} \right ] \qquad \b{ \textbf{Y} = \left [ \begin{matrix} 0 & -i \\ i & 0 \end{matrix} \right ]\qquad \textbf{Z} = \left [ \begin{matrix} 1 & 0 \\ 0 & -1 \end{matrix} \right ]}
\label{eq:matrices}\end{equation}
\b{Single qubit gates represent rotations on the Bloch sphere. }Performing a gate is logically equivalent to multiplying the ket column vector by the matrix of the gate. \b{The \textbf{I} gate leaves the qubit unmodified.} The \textbf{X} gate \b{rotates the qubit by $\pi$ around the X-axis}, which flips the amplitudes for the $\ket 0$ and $\ket 1$ states. \b{Similarly, the \textbf{Z} gate rotates the qubit around the Z-axis and the \textbf{Y} gate rotates the qubit around the Y-axis.}  Other common gates include the Hadamard (\textbf{H}) gate and phase gates \textbf{S} and \textbf{T}, \b{shown in Equation} \ref{eq:gates2}.
\begin{equation}
\b{\textbf{H} = \frac{1}{\sqrt 2}\left [ \begin{matrix} 1 & 1 \\ 1 & -1 \end{matrix} \right ]}\qquad
\b{\textbf{S} = \left [ \begin{matrix} 1 & 0 \\ 0 & i \end{matrix} \right ]}\qquad
\b{\textbf{T} = \left [ \begin{matrix} 1 & 0 \\ 0 & e^{i\pi /4} \end{matrix} \right ]}
\label{eq:gates2}
\end{equation}
\b{The \textbf{S} gate is a rotation around the Z-axis by $\pi/2$ and the \textbf{T} gate is a rotation around the Z-axis by $\pi/4$. Performing two \textbf{T} gates back to back is equivalent to an \textbf{S} gate, and two \textbf{S} gates back to back is equivalent to a \textbf{Z} gate. The \textbf{H} gate is commonly used at the beginning of quantum algorithms to put qubits into a perfect superposition state.}

Two-qubit gates commonly involve a control qubit and a target qubit. In this case, a gate is performed on the target qubit if the control qubit is in the $\ket 1$ state. For example, the controlled-NOT, \textbf{CNOT} gate is a controlled-\textbf{X} gate. \b{Two-qubit gates are used to create entanglement between the qubits.}

Quantum gates must be \emph{unitary}. This means they must be linear, reversible, and preserve the magnitude of the column vector. E.g., the \textbf{X} gate is its own inverse, and if two \textbf{X} gates are applied sequentially, the qubit returns to the original state. In other words, unitary operations coherently transform the quantum state. Conversely, measurements are non-unitary and irreversible. If a qubit in a superposition of $\ket 0$ and $\ket 1$ is measured and found to be $\ket 0$, it is then entirely in the state $\ket 0$. This is an incoherent process, as the quantum state has effectively been destroyed, containing only classical information. Whether operations are unitary or not is important not only for quantum gates and measurements, but also for the noise that affects the quantum state.

Quantum states that are not pure are called \emph{mixed states}. These states are combinations of pure states, each with an associated classical probability. Mixed states occur as the result of imperfect isolation and manipulation of the quantum state, which applies to all physically realizable quantum states. \emph{Density matrices} are the equivalent of kets for mixed states. Density matrices can represent all pure and mixed states. A ket representation can be converted to a density matrix by taking the outer product of the ket with its conjugate transpose (adjoint)
\begin{equation} \left [ \begin{matrix} \alpha \\ \beta \end{matrix} \right ] \longrightarrow \left [ \begin{matrix} \alpha \alpha^* & \alpha \beta^* \\ \beta \alpha^* & \beta \beta^* \end{matrix} \right ] \end{equation}
where $^*$ denotes the complex-conjugate. 
A common metric to quantify the ``quality'' of a quantum state is the fidelity \cite{flammia2011direct}, which is defined as
\begin{equation}
    Tr[ \rho \sigma ]
\end{equation}
where the density matrix $\rho$ represents the ``correct'' quantum state; $\sigma$, the actual quantum state; and Tr, the trace (diagonal sum) of the matrices multiplied. 
%
%If we have desired, ``correct'' quantum state, represented by the density matrix $\rho$, and an actual state, represented by $\sigma$, we can define the fidelity \cite{flammia2011direct} of the actual state as 
%\begin{equation}
%    Tr[ \rho \sigma ]
%\end{equation}
%which is the trace (diagonal sum) of the matrices multiplied. This gives us a metric to quantify the ``quality'' of quantum state. 
%
Simulating with density matrices allows one to keep track of classical probabilities and possible errors, in addition to the quantum transformations. This can be convenient in many cases \cite{densitymatrixgutierrez2016errors}, however the  computational resources required increase significantly \cite{beale2018quantum}.
\section{Quantum Noise}
\label{sec:noise}
Noise is present in all computing systems. However, it is quite a force to be reckoned with for quantum systems. In fact, noise is so pervasive that it is impossible to have a meaningful discussion about practical quantum computing without an in-depth consideration of its effects. Clearly, no benchmarking approach can succeed without considering noise and the resulting impact on measured or simulated results. This has been unfortunate, as quantum noise is difficult to characterize and, in many cases, its effects are not well understood. Here, we provide a brief overview of quantum noise and the models used to represent it.

\subsection{Physical Sources of Noise}
Quantum noise can come from a variety of sources. There is never only a single source of noise in any given quantum system. Determining what the sources are and what their relative contributions are is a hard problem.  Possible sources are unwanted interaction with the environment (both distinct events and inevitable decay of quantum states), unwanted interaction between qubits, or imperfect control operations. Each of these introduce error with significantly different characteristics, giving rise to 
%and there are
different models.
%for each kind of noise. 
Here, we go over different common sources and discuss their physical impact. A summary of physical noise sources is provided in Table \ref{tab:physicalnoise}.

%\vspace{-.1cm}
\subsubsection{Interaction with the Environment}
\label{sec:intwenvironment}
Qubits need to be perfectly isolated from the environment to maintain their state. If such a system could be constructed, there would be no quantum noise. But no real system can be perfect, hence there is inevitably some interaction. This can be seen as a ``measurement'' of the system \cite{barnes2017quantum}, as information is leaving the quantum state. As measurements are non-unitary, this kind of noise is also non-unitary. The expected amount of time a system can remain unperturbed is called the coherence time. Commonly reported are the $T_1$ and $T_2$ times. $T_1$ measures the expected loss of energy from the system; if a qubit is put into an excited $\ket 1$ state, $T_1$ is a measure of how long it takes to collapse to the $\ket 0$ state. \b{This is also called the qubit relaxation time \cite{tomita2014low}}. $T_2$ measures the dephasing time; if a qubit is placed in the superposition state $\ket 0 + \ket 1$, $T_2$ determines how long it takes to polarize either to $\ket 0$ or $\ket 1$ \cite{nielsen2002quantum}. Risk of interaction with the environment is increased when performing operations on the qubits, as the driving force of the operation comes from an external input. {This is an unfortunate situation as two critical requirements have conflicting needs. The quantum state needs near perfect isolation to remain intact, yet also must interact with control mechanisms in order to perform useful computation. This is referred to as the coherence-controllability trade-off \cite{yoneda2018quantum}.}

Interaction with the environment can also produce unitary errors, such as global external fields which act on the qubits \cite{greenbaum2017modelingcoherent,vgt,vgt2,vgt3,wallman2015estimating}. \b{Such interactions can cause unitary rotations of the quantum state.}
%\redHL{Intuition? Also add to this section; that fundamentally quantum computing translates into anti-isolation.}

%\vspace{-.1cm}
\subsubsection{Interaction with Other Qubits}
As previously mentioned, qubits can become entangled with each other. This means their states become correlated. While this is a frequently used tool in quantum computation, it needs to occur only when desired. If qubits interact accidentally, this can lead to a mixture of their quantum states or decoherence \cite{brecht2016multilayer,pratt2001qubit,sarovar2019detecting}. This is referred to as cross-talk. This type of error has been particular difficult to characterize. 

If left in perfect isolation, cross-talk between qubits would lead to a unitary evolution of the state. Hence, all the information is still contained in the quantum state. However, the quantum state would be different than the one desired, which destroys the ability to manipulate it in a meaningful way. For example, we may wish to have two qubits that are far apart 
%that are 
and highly entangled, in order to perform quantum teleportation \footnote{Quantum teleportation is the transfer of quantum information between qubits by means of quantum entanglement in combination with a classical channel. It can be used to transfer information between non-adjacent qubits in a quantum computer or over long distances via a quantum network.}. However, if they interact with other, nearby qubits, this will disperse and decay the entanglement \cite{pratt2001qubit}.  If not in perfect isolation, cross-talk can cause increased degradation of the quantum state. Say one is performing error correction, which involves interacting extra (ancilla) qubits with the qubits that store the data, and then measuring the ancilla to extract error information (syndromes). 
Cross-talk may cause the unintentional interaction of a data qubit with an ancilla qubit. Thereby some of the quantum information in the data qubit can get transferred into the ancilla. As a result, when the ancilla is measured, the computer would unknowingly extract information from the data qubit, corrupting its quantum state.
%
%If there is cross-talk, the ancilla qubit may have unintentionally interacted with a data qubit it %wasn't supposed to. Hence, some of the quantum information in this data qubit was transferred into the ancilla. When the ancilla is measured, the computer unknowingly extracts information from the data qubit, corrupting its state.

\subsubsection{Imperfect Operations}
Imperfect application of quantum gates can generate incorrect quantum states. Often, these are slight over- or under- rotations which are the result of imperfect calibration \cite{bravyi2018correctingcoherenterrors}. These kinds of errors do not directly destroy the quantum state but lead to coherent evolution of the quantum state into an undesired state \cite{barnes2017quantum}. This type of noise is predominantly seen in modern experiments \cite{wallman2016noise} and its potentially catastrophic impact on the ability to perform error correction has been a concern in recent years \cite{bravyi2018correctingcoherenterrors}.

\subsubsection{Leakage}
Many quantum systems that are used as qubits actually have more than two possible states. In this case, two of the possible states are selected to represent $\ket 0$ and $\ket 1$. \b{In other words, the qubit is encoded in a subspace of a larger quantum system \cite{wood2018quantification}. This subspace is called the \emph{computational subspace}.} It is assumed that the quantum systems remain in these two states (though other states may be used temporarily, such as in the implementation of two-qubit gates \cite{bruzewicz2019trapped}). If a qubit unintentionally enters one of these other states, it is referred to as leakage, and it can be particularly detrimental \cite{fowler2013coping}. \b{The return of a qubit back into the compuational subspace is called \emph{seepage} \cite{wood2018quantification}. } \b{Leakage and seepage can be either unitary or non-unitary, and can be caused by imperfect control or unwanted interactions with the environment \cite{wood2018quantification}.}

\begin{table}[h]
%    \scalebox{.91}{
    \centering
    \begin{tabular}{c||c|c}
& Environment & Other Qubits \\
\hline \hline
 Unitary & \begin{tabular}{@{}c@{}}External Fields \\ Over/under Rotations from Imperfect Control \cite{barnes2017quantum} \\  \end{tabular} & Cross-talk \cite{pratt2001qubit,brecht2016multilayer}\\
 \hline  
Non-Unitary & Unintentional Measurements \cite{barnes2017quantum} \\   & \\
    \end{tabular}
%    }
   \vspace{.1cm}
    \caption{Categorization of physical noise into its sources and whether it is unitary or not.}
%    \vspace{.7cm}
    \label{tab:physicalnoise}
\end{table}

\subsection{Noise Models} 
%\redHL{(maybe another Table tying this Sect. to the prev.?)}} 
%If one is working slightly higher in the system stack, 
Working at higher levels of the system stack, 
%it may not be of interest where the source of noise is. However, 
it is more critical to know how quantum noise will affect quantum operations. In this section, we transfer focus from physical sources of noise to the process of modeling them. The goal is to learn how noise disrupts the correctness of quantum algorithms during operation. As most research labs do not have their own physical quantum computer, and publicly available machines have low qubit counts, accurate and efficient noise models are greatly desired to facilitate simulation. Quantum noise is notoriously difficult to model accurately \cite{wallman2016noise}. There are variations of quantum noise and it is possible that it may even be non-Markovian.
%\cite{} \redHL{(definition + missing citation)}. 
Knowing what specific type of noise is present and how it affects a particular system is difficult to determine without extensive, physical experiments. However, there are a number of possible methods to estimate noise, with varying degrees of accuracy and computational efficiency. Realistic noise models are often intractable to simulate at scale, so simplifying assumptions are made to reduce complexity \cite{magesan2012efficient}. It is important to know when these assumptions are appropriate to make in order to produce realistic results. 
%Some 
%Most models are only appropriate under 
%certain 
%strict
%conditions. 
Here, we give a brief, high-level overview of different noise models and discuss their implications. A summary of noise models is provided in Table \ref{tab:noisemodels} and a summary of the physical noise processes they emulate is shown in Table \ref{tab:physical2model}.

\subsubsection{Stochastic Pauli Noise}
{Stochastic Pauli noise is the simplest and most intuitive noise model. Additionally, according to the well-known Gottesman-Knill theorem \cite{gottesman1998heisenberg,aaronson2004improved}, it is easy to simulate using classical computers \cite{bravyi2018correctingcoherenterrors} and, at the same time, easy to correct using standard error correction procedures \cite{wallman2016noise}.} 
%\redHL{intuition? footnote? or no need?} 
Hence, it has become popular \cite{li2019sanq,Qiskit,khammassi2017qx}. It is most applicable for modeling unwanted interactions with the environment, which is effectively unintentional ``measurements'' of the quantum state \cite{barnes2017quantum}. It can be implemented by inserting an \textbf{X}, \textbf{Y}, or \textbf{Z} gate into a circuit at random with some specified probability. The effect on the overall fidelity can be estimated with Monte Carlo simulation \cite{li2019sanq}. Alternatively, representing the quantum state as a density matrix, $\rho$, the noise can be modeled as 
\begin{equation}\label{eq:dmpauli}
 N_i(\rho) = (1-\epsilon_i)\rho + \epsilon_i^x \textbf{X}\rho \textbf{X} + \epsilon_i^y \textbf{Y} \rho \textbf{Y} + \epsilon_i^z \textbf{Z} \rho \textbf{Z} 
\end{equation}
where $\epsilon_i$ is the total error rate on qubit $i$ and $\epsilon_i^x$, $\epsilon_i^y$, and $\epsilon_i^z$ are the rates for each type of error, corresponding to the probabilities of inserting each gate \cite{bravyi2018correctingcoherenterrors}. \b{This is also referred to as \emph{depolarizing noise} \cite{tomita2014low}. If $\epsilon_i^x = \epsilon_i^y = \epsilon_i^z$, it is called \emph{symmetric depolarizing noise}.} $\textbf{X}$, $\textbf{Y}$, and $\textbf{Z}$ are operators performing the respective gate on the qubit.{ While \textbf{X}, \textbf{Y}, and \textbf{Z} gates are unitary operations (causing a coherent transformation of the quantum state), inserting them in a probabilistic manner does not represent a coherent process. Additionally, the linear combination of unitary operations, as in Equation \ref{eq:dmpauli}, can represent a non-unitary operation. Hence, stochastic Pauli noise is an incoherent source \cite{bravyi2018correctingcoherenterrors}.} 

A common strategy is to inject error only after each gate. However, this is not realistic as qubits can acquire error even when remaining idle \cite{WallmanISCA}. Therefore, Pauli noise should be injected in every cycle. Numerous studies have found that stochastic Pauli noise models often lead to inaccurate and overly optimistic results \cite{nickerson2019analysing,bravyi2018correctingcoherenterrors,greenbaum2017modelingcoherent,gutierrez2015comparison,gutierrez2013approximation,barnes2017quantum}, but that they still can provide reasonable approximations in 
%some 
certain
conditions. These include errors at the logical level under QEC \cite{bravyi2018correctingcoherenterrors,greenbaum2017modelingcoherent,gutierrez2015comparison,beale2018quantum}. 

There are some natural extensions to this model which can result in more accurate simulations. Significant improvements can be made, while remaining efficiently simulable, by augmenting Pauli gates
%a Pauli channel 
with Clifford group operators (i.e., Hadamard (\textbf{H}), phase (\textbf{S}), and \textbf{CNOT} gates) and Pauli measurements \cite{gutierrez2013approximation}. 
%\r{what is a channel? what are Clifford ops?} 
{This involves the same process of inserting gates at random, but using a larger gate set, which,
%This larger gate set has, 
in addition to the Pauli gates (\textbf{I},\textbf{X},\textbf{Y},\textbf{Z}), has \textbf{H}, \textbf{S}, and \textbf{CNOT} gates \cite{magesan2013modeling}.} %\redHL{(all has to be defined)} 

A fundamental problem with stochastic Pauli noise is that it is ``not quantum enough'' \cite{bravyi2018correctingcoherenterrors}. While the inserted Pauli gates are quantum operations, the choice of whether to insert them is based on a classical probability. While a classical noise model is familiar and intuitive from a computer architecture perspective, it is not necessarily true depiction of the real errors occurring at the physical level. 

\subsubsection{Coherent Noise}
Coherent noise models attempt to capture evolution of the quantum state that, while not destructive, is still undesired. One could see this as coherently performing a quantum program, just one that is different than intended. Physical coherent noise can be caused by imprecise classical control of the quantum operations \cite{bravyi2018correctingcoherenterrors}, external fields, and cross-talk \cite{greenbaum2017modelingcoherent}. Modeling coherent noise can be difficult as some of the relevant sources of noise are not well understood. Hence, simplifying assumptions are made. While not exact, the goal is for the model to affect the quantum state similar to realistic sources. Examples include static Z-rotations \cite{bravyi2018correctingcoherenterrors}, X-rotations in combination with Pauli-X errors \cite{greenbaum2017modelingcoherent}, or rotations about a non-Pauli axis \cite{gutierrez2013approximation}. Coherent noise is not efficiently simulable, meaning the classical resources required to simulate grow exponentially with the system size. Hence, these simulations are limited to relatively small systems \cite{li2017fault,tomita2014low,barnes2017quantum,chamberland2017hard}.

Coherent noise is typically much more detrimental, with a much higher worst case error rate \cite{wallman2016noise,bravyi2018correctingcoherenterrors,barnes2017quantum}. Additionally, {many quantum algorithms consist of periodic circuits, where the same sequences of gates are repeated many times. Coherent noise is particularly harmful in this case, where its effects get amplified with each iteration \cite{blume2019volumetric,blume2017demonstration}.} From this, it may appear that coherent noise is more important, and should be assumed unless known otherwise. However, this may not be true for all circumstances. The effects of coherent noise were analyzed on quantum error correction \cite{greenbaum2017modelingcoherent,bravyi2018correctingcoherenterrors}. It was found that the coherence of the error is reduced at the logical level, and is further decreased with a higher code distance. This means that it may be sufficient to assume stochastic Pauli noise at the logical level, even if the physical noise is coherent. However, it was noted that using a stochastic Pauli model for physical noise would significantly under estimate the error rate at the logical level.

Unfortunately, as modern quantum computers are not capable of QEC, they cannot make use of this resilience to coherent error. Randomized Compiling (RC) \cite{wallman2016noise} is a novel approach which may help in this domain. {The basic idea is to perform randomizing Pauli gates during the run of a quantum circuit, which are interleaved with the gates of the program. At each location that randomization is introduced, the previous randomization is undone to return the quantum state to the desired state. These randomizing }operations disrupt the coherent noise and tailor it effectively into stochastic Pauli noise. { Prior to execution, these additional randomizing gates can be fused with (compiled ``into'') the actual gates in the circuit. Depending on the gate set available in the system, RC 
%may be able to 
can be performed with no overhead. }
%compiled into the circuit and therefore do not necessarily correspond to any additional operations, 

\begin{table}[h]
%\scalebox{0.91}{
    \centering
    \begin{tabular}{c||c|c}
 & Efficient & Not Efficient \\
\hline \hline
 Unitary & NA & \begin{tabular}{@{}c@{}} Coherent Rotations \cite{bravyi2018correctingcoherenterrors} \\  \end{tabular}  \\
 \hline
Non-Unitary &  \begin{tabular}{@{}c@{}}Stochastic Pauli \\ Pauli Twirling \cite{Paulitwirling} \\ Clifford Channels \cite{gutierrez2015comparison,gutierrez2013approximation} \end{tabular} & Amplitude/Phase Damping\cite{tomita2014low}
    \end{tabular}
%    }
    \vspace{.1cm}
    \caption{Categorization of noise models into  whether they are unitary and whether they are efficiently simulable or not.}
    \label{tab:noisemodels}
\end{table}

%\vshrink{.5}
\subsubsection{Amplitude/Phase Damping}
Amplitude Damping (AD) is a non-coherent error model which captures energy loss from the quantum state into the environment, such as spontaneous emission of a photon \cite{tomita2014low}. This noise model is relevant to any quantum system with multiple energy levels, where there is an excited state and ground state, with a tendency for the excited state to decay to the ground state,  
%\redHL{(missing citation + what is a ``degenerate'' energy level? define)}, 
such as ion-trap quantum computers which use the excitation levels of electrons. Additionally, the loss of energy must be to some environment -- i.e., in the previous example, if the energy loss is a spontaneous emission of a photon, there must be an environment for the photon to escape into. Hence, if the quantum system was \emph{perfectly} isolated, AD would not occur. AD is a realistic noise model but is also not efficiently simulable. However, models have been designed to approximate the effect of AD, but which remain simulable \cite{gutierrez2013approximation}, such as Pauli Twirling \cite{PTsilva2008scalable,PTghosh2012surface,PTsarvepalli2009asymmetric,tomita2014low}. \b{Phase damping (PD), sometimes call pure dephasing, is equivalent to a phase flip channel \cite{tomita2014low}. This does not change the probability of a qubit being in either the $\ket 0$ or $\ket 1$ state, but it changes the phase between the two states. AD is closely related to the $T_1$ time and PD is closely related to the $T_2$ time, which are discussed in Section \ref{sec:intwenvironment}}.%\redHL{(incomplete sentence?)}

%\subsubsection{Other Non-Unitary Noise}
%There are other sources of error that affect the quantum state that are non-unitary, such as amplitude damping.
%\redHL{(merge w/ previous sect.?)}. 
%Such noise deforms the Bloch sphere \cite{puzzuoli2014tractable}. Such noise can be modelled more accurately at the physical level with Clifford Channels, but it was found that when using Steane error correction, Pauli Channels produce accurate estimates \cite{gutierrez2015comparison}. 
%\redHL{What is Clifford channel, Steane EC or Pauli Channel? Need also brief primer on QEC/noise modeling as it appears...}

%\subsection{Take Away}

\vspace{.1cm}
\noindent\boxed{
\begin{minipage}[t]{0.99\columnwidth}
%Takeaway: The lesson is that
The nature of quantum noise greatly affects the quantum state, and by direct result, the performance of any potential quantum computer. Hence, even at higher levels in the system stack, one must give serious attention to the expected noise present in the system and be sure it is adequately accounted for. This impact of quantum noise is often overlooked or given secondary consideration. In many cases stochastic Pauli noise may be an overly simplified model. If so, it will produce incorrectly optimistic results, especially for modern systems.
\end{minipage}
}
%\vspace{-.2}

%\redHL{frame it, and make the message crisper/more concrete}

\begin{table}
%\scalebox{.91}{
    \centering
    \begin{tabular}{c|c}
        Physical Noise Source & Noise Models \\
        \hline
        \hline
         Interaction with Environment & \begin{tabular}{@{}c@{}}Stochastic Pauli Noise \\ Amplitude/Phase Damping \\ Pauli Measurements  \end{tabular} \\
         \hline
         Imperfect Control & Coherent Over/Under Rotations\\
         \hline
    \end{tabular}
%    }
    \vspace{.1cm}
    \caption{Physical noise and corresponding noise models}
    \label{tab:physical2model}
\end{table}

%\begin{tabular}{@{}c@{}} \\  \\   \end{tabular}

%Additionally, the fidelity of single- or two-qubit gates in isolation do not translate well to the process fidelity of the entire system. Hence, experiments demonstrating exceedingly high gate fidelity are not as encouraging as they may seem.

%New subsection on metrics
\section{Metrics}
\label{sec:metrics}
Metrics enable quantitative analysis on the quality of a given system. Choosing metrics can be tricky, and can be misleading if not done wisely. This is true for classical systems and, unsurprisingly, is even more complicated for quantum systems. A number of quantum metrics have been developed and used widely, however there is no unifying ``gold standard'' \cite{gilchrist2005distance}. Depending on the use case, different metrics may be preferable.   

A commonly used metric for quantum operations is the \emph{process fidelity}. The process fidelity of a noisy quantum operation, $\tilde G$, relative to the noiseless operation, $G$, can be determined by \cite{erhard2019characterizing,flammia2011direct}
\begin{equation} F(G(\rho ), \tilde G(\rho)) = Tr \left[ G(\rho ) \tilde G(\rho) \right ]
\label{eq:fidelity}
\end{equation}
where $\rho$ is the density matrix representing the input quantum state and {$Tr$ is the trace. At a high level, process fidelity measures ``how similar'' the noisy output quantum state is to the intended target. If the noisy operation $\tilde G$ is free of error, the process fidelity will be 1.} One drawback is that it is not strongly tied to a physical interpretation, when comparing two different mixed states \cite{gilchrist2005distance}. However, if one of the states is a pure state (represented here by $G(\rho)$), and the other is a noisy mixed state (represented here by $\tilde G(\rho)$), then it is the overlap of the two states \cite{gilchrist2005distance}.  

Another commonly used metric is the \emph{average gate fidelity}, or conversely the \emph{infidelity}, the average gate infidelity to the identity \cite{beale2018quantum}. The fidelity of a quantum gate is its process fidelity, where the qubit's quantum state after applying the gate is compared to what it should be. This is complicated by the fact that the qubit could be in any possible state prior to the operation (anywhere on the surface of the Bloch Sphere in Figure \ref{fig:blochsphere}) and the fidelity could be different for each state. A general solution to this is to average over all possible input states, i.e., to integrate over the surface of the Bloch Sphere \cite{cabrera2007average,bowdrey2002fidelity}:
\begin{equation}
    \langle F \rangle = \frac 1 {4\pi} \int Tr\left [G(\rho)\tilde G(\rho) \right ] d \Omega
\end{equation}
\noindent However, it was shown that it is also possible to achieve the same result by averaging over a finite set of inputs \cite{bowdrey2002fidelity}. 

The \emph{trace distance} \cite{ruskai1994beyond} is an alternative which measures the distinguishability of two quantum states. It is defined as 
\begin{equation}
    D(G(\rho),\tilde G(\rho))= \frac{1}{2} Tr \left[ \sqrt{(G(\rho)-\tilde G(\rho))^{\dagger} \; (G(\rho)-\tilde G(\rho))}\right ]
    \label{eq:tracedistance}
\end{equation}
\noindent where again $\rho$ is the density matrix for the input quantum state. \b{ $G(\rho)$ is density matrix of the ideal output quantum state, which is $\rho$ transformed by the process $G$. $\tilde G(\rho)$ is the density matrix of the noisy output quantum state, which is $\rho$ transformed by the noisy process $\tilde G$.} $\dagger$ means the adjoint. The trace distance is related to the process fidelity via
\begin{equation}
    1 - \sqrt{F(G(\rho),\tilde G(\rho))} \leq D(G(\rho),\tilde G(\rho)) \le \sqrt{1-F(G(\rho),\tilde G(\rho))}
\end{equation}
allowing bounds to be placed on one if the other is known. The diamond distance \cite{fuchs1999cryptographic} is the maximum trace distance between any state and the state impacted by noise \cite{blume2017distinguishable} 
\begin{equation}
    ||G - \tilde G || _d = max_\rho \left ( || ( G \otimes \textbf{I})[\rho] - (\tilde G \otimes \textbf{I})[\rho]||_1 \right )
    \label{eq:diamond}
\end{equation}
where \cite{wallman2016noise}
\begin{equation}
    ||M||_p = \left ( Tr(M^\dagger M)^{p/2} \right ) ^{1/p}
\end{equation}
is the Schatten p-norm of M. Note that when $p=1$, Equation \ref{eq:diamond} takes a form very similar to Equation \ref{eq:tracedistance}. The diamond distance is a measure of the distinguishability of $G$ and $\tilde G$ given a single use of either operator. This will be quite small for any well-performing hardware \cite{blume2017distinguishable}. Here, $\rho$ can be entangled with additional ancilla qubits, and these inputs are often the most sensitive \cite{blume2017distinguishable}. The diamond distance is often used in proving fault-tolerance of quantum computers when doing rigorous analyses of errors in quantum circuits \cite{shor1995scheme,aharonov2008fault,knill1998resilient} as it represents the worst-case error. A drawback of this is that it may be overly pessimistic \cite{wallman2015error}. While there are no efficient known ways of computing it directly, there are methods of efficiently computing the bounds \cite{wallman2015error}. 

If one is interested in only the output measurement probabilities, the \emph{Hellinger fidelity} may be a more appropriate metric. The Hellinger fidelity is defined as $1 - Hellinger \;distance$. The Hellinger distance measures difference between two probability distributions \cite{luo2004informational}. For example, if we have two probability distributions $p$ and $q$, and $p_i$ and $q_i$ are the probability of sampling $i$ in each distribution, the Hellinger distance is defined as
\begin{equation} \frac {1} {\sqrt{2}}  \sqrt{\sum _i (\sqrt p_i - \sqrt q_i )^2 } \end{equation}
If the distributions are identical, the Hellinger distance is 0, and if there is no overlap, the Hellinger distance is 1. The quality of a noisy quantum operation is quantified by determining the Hellinger fidelity of its measurement distribution to that of an ideal one. This is convenient in experiment, as all that is required is repeated iterations of the noisy operation and measurements in the standard Z-basis. It is provided as a standard metric on IBM's Qiskit \cite{aleksandrowicz2019qiskit}.

Many other variants of distance metrics exist \cite{spehner2017geometric,luo2004informational}. We have provided introductory definitions to some  widely-used metrics, but refer the interested reader to  comprehensive works for more in-depth discussion on which quantum metrics are useful and in which contexts \cite{gilchrist2005distance}.

\section{Qubit Benchmarking}
\label{sec:qubit}
%Quantum computers use quantum bits (qubits) in place of bits. Qubits can exist in two states, corresponding to logical 1 and 0. However, they can exist in both at the same time (\textit{superposition}) and can be dependent on each other (\textit{entangled}). These properties enable quantum computers to perform algorithms that will never be possible on classical computers. We do not discuss these principles but refer the interested reader to an excellent introduction in \cite{loceff2015course}. 

%Quantum operations performed on qubits are called quantum gates. Qubit states can be visualized as vectors on the Bloch sphere and single-qubit quantum gates are represented by rotations around the different axes. Common examples are the X, Y, Z, and H(adamard) gates. There are also multi-qubit gates, such as the CNOT gate where the state of a target qubit is flipped if the control qubit is 1. Not all quantum hardware can implement every type of gate. However, individual quantum gates can be replaced with sequences of alternate gates which implement the same functionality. Therefore, a quantum computer needs only to have a \emph{universal} set of quantum gates. This is equivalent to a universal gate set for classical computers, eg. NAND. 

Before talking about benchmarking a quantum computer, we have to talk about benchmarking the qubits themselves, even in a single (or two) qubit system. If one is constructing a quantum computer, it is clearly of great interest how reliable the quantum operations (gates) are. The average gate fidelity coarsely predicts how many gates can be applied before the quantum state gets too corrupted. Additionally, quantum error correcting codes only work if the error is below a certain threshold \cite{knill1998resilient}. On top of this, the overhead of the error correction strongly depends on the fidelity \cite{oskin2002practical}. Quantum gates suffer from high error rates. Errors in classical switches could be less than 1 in $10^{15}$, 
%\redHL{of what?}, 
whereas effective quantum error rates are frequently above $1\%$, where the previously mentioned error sources and models apply and determining their impact is a complex task.

Unfortunately, experimentally determining quantum states, and the fidelity of quantum operations, is not straight forward. Measurements of a quantum system are destructive, so the state will need to be prepared or the operation performed for each measurement. In addition to this time overhead, errors in the initial state preparation and the measurements, (SPAM) errors, can obscure the error in the operation. 

A brute force approach to learn a quantum state is quantum state tomography \cite{vogel1989determination}. This enables the classical extraction of all quantum information, and hence answers any questions we may have about its structure. This requires measuring a complete set of observables (physical properties that can be measured) which determines the quantum state. For example, say there is a single qubit in the quantum state 
\begin{equation} \ket \psi  = \sqrt{0.8}\ket 0 + \sqrt{0.2}\ket 1 \end{equation}
For numerical clarity we use a pure state, however this process works for mixed states, as well. While this is a pure state, it can be described equivalently in density matrix form: 
\begin{equation} \rho = \left [ \begin{matrix} 0.8 & 0.4 \\ 0.4 & 0.2 \end{matrix} \right ] \end{equation}
\noindent This state is unknown to the outside world, but the state can reliably be prepared by a quantum operation. The goal then is to determine this density matrix only by performing repeated measurements on the prepared state $\ket \psi$. All single qubit density matrices can be represented as a linear sum of the Pauli matrices \cite{paris2004quantum}:
\begin{equation}
    \rho = \frac 1 2 \left (S_0 \textbf{I} + S_1 \textbf{X} + S_2 \textbf{Y} + S_3 \textbf{Z} \right )
\end{equation}
\noindent Therefore, if we find the coefficients, we can reconstruct the density matrix. These coefficients can be found by determining measurement probabilities when measuring along the X, Y, and Z bases \cite{paris2004quantum}
\begin{equation}
\begin{aligned}
    S_0 &= P_{\ket 0} + P_{\ket 1} &= 0.8 + 0.2 &= 1.0\\
    S_1 &= P_{\ket 0 + \ket 1} - P_{\ket 0 - \ket 1} &= 0.9 - 0.1 &= 0.8\\
    S_2 &= P_{\ket 0 + i\ket 1} - P_{\ket 0 - i\ket 1} &= 0.5 - 0.5 &= 0.0\\
    S_3 &= P_{\ket 0} - P_{\ket 1} &= 0.8 - 0.2 &=0.6\\
\end{aligned}
\label{eq:coefficients}
\end{equation}
\noindent where $P_{\ket \phi}$ is the probability of measuring $\ket \phi$. Note that $S_0 = 1$ by construction, as $\ket 0$ and $\ket 1$ are the only two measurement outcome possibilities. $S_3$ only requires measurements along the standard Z-basis. To find $S_1$ and $S_2$, measurements need to be performed along the X and Y axes. This can be done with a basis change prior to measurement, which is done by applying the appropriate quantum gate just prior to measurement. For example, performing a Hadamard gate then a Z-basis measurement is effectively an X-basis measurement. Note that each measurement must be performed many times to achieve accurate probability estimates. The exact probabilities are shown in Equation \ref{eq:coefficients}. For the example we computed them directly with $Tr\left [\xi \rho  \right ]$, where $\xi$ is the density matrix of the corresponding basis state. Once the coefficients are known the state can be described
\begin{equation} 
\rho = \frac 1 2 \left ( 1 * \left [ \begin{matrix} 1 & 0 \\ 0 & 1 \end{matrix} \right ] + 
0.8 * \left [ \begin{matrix} 0 & 1 \\ 1 & 0 \end{matrix} \right ] + 
0 * \left [ \begin{matrix} 0 & -i \\ i & 0 \end{matrix} \right ] + 
0.6 * \left [ \begin{matrix} 1 & 0 \\ 0 & -1 \end{matrix} \right ] \right ) =
\left [ \begin{matrix} 0.8 & 0.4 \\ 0.4 & 0.2 \end{matrix} \right ]
\end{equation}
This process can be generalized to multi-qubit systems as well, we refer the reader to a thorough introduction in \cite{paris2004quantum}. However, it does not scale well as it requires a number of measurements which is exponential in the system size. In addition to this, it is difficult to distinguish states with low probability from those with zero \cite{knill2008randomized} and the computation to convert measured results into an estimate of the quantum state is intractable \cite{cramer2010efficient}. Hence, this approach is not feasible for large systems. There are notable variations of this process, such as Shadow Tomography \cite{aaronson2020shadow}. While quantum state tomography has exponential cost because it tries to answer all possible questions, shadow tomography attempts to only learn certain features by learning from measurements \cite{huang2020predicting}. The name comes from the idea that one is not trying to learn the full density matrix, only the ``shadow'' it casts on the chosen measurements \cite{aaronson2020shadow}.

Quantum process tomography (QPT) uses the same approach as quantum state tomography, except that the goal is to identify a quantum operation, rather than the quantum state. Known input quantum states are generated and then the operation is performed. Quantum state tomography is then applied to the output states allowing identification of the process \cite{chuang1997prescription,poyatos1997complete}. As it follows the same procedure as quantum state tomography, it also requires exponential resources in the number of qubits. The key assumption that QPT makes is that the input state is known, and therefore only the quantum operation needs to be found. This simplifies the problem, and finding the parameters of the operation is equivalent to maximum likelihood estimation of a convex objective function (which has a single, global minimum) \cite{greenbaum2015introduction}. Hence, standard convex optimization techniques can be used to solve for it \cite{chow2012universal,boyd2004convex}. This simplification comes at a cost however. The input state isn't necessarily known, as there can be errors in the state preparation process, which can lead to inaccuracy \cite{greenbaum2015introduction}.  Compressed quantum process tomography \cite{shabani2011efficient} uses compressed sensing, a known classical signal processing strategy, to reduce the number of measurements required. The unitary operations performed by quantum computers can typically be represented by nearly-sparse process matrices, and hence can be well approximated by sparse matrices. These sparse matrices can be found with exponentially fewer measurements \cite{shabani2011efficient}.

Direct Fidelity Estimation \cite{flammia2011direct} is another clever method used to extract meaningful information without exponential overhead. This procedure estimates the fidelity of an arbitrary quantum state relative to a pure state. Here, the pure state is an error free state which is the intended result of a quantum operation; and the arbitrary state, what is actually produced. Note that complexity can be reduced by not attempting to learn everything about the state, seeking only to get the estimate of the fidelity of the arbitrary state. The estimate is achieved by measuring a constant number of Pauli observables (applying standard Pauli operations and then measuring). This is possible by making educated guesses about which Pauli observables are likely to reveal errors.   

Gate set tomography (GST) is an extension of QPT \cite{greenbaum2015introduction,nielsen2020gate,merkel2013self}. Like QPT, the resources it requires increase exponentially with the number of qubits \cite{greenbaum2015introduction}. Hence, it cannot target large quantum systems and is also used mainly for 1- and 2-qubit systems. However, a significant advantage it has over QPT is that it is calibration-free, i.e., it does not depend on accurate descriptions of the initial prepared states \cite{nielsen2020gate}. This is significant, as QPT can generate highly inaccurate results when the gates used to prepare the input states have systematic error \cite{merkel2013self}. Like QPT, the process of finding the parameters of the quantum operations (gate set) involve maximum likelihood estimation of an objective function based on measurement results. However, the inclusion of state preparation and measurement (SPAM) errors in the gate set produces a non-convex objective function \cite{greenbaum2015introduction}. Hence, standard convex optimization techniques do not work, and a combination of approximate and iterative methods must be used. Another consequence of including SPAM errors is that it is not possible to characterize a single gate at a time, as in QPT. Rather, there is a minimal set of gates which must be estimated simultaneously \cite{greenbaum2015introduction}. The mathematical background of GST is provided in \cite{nielsen2020gate} and the protocol is provided in \cite{greenbaum2015introduction}.

A scalable approach which has been utilized frequently in recent years is \textit{Randomized Benchmarking} \cite{knill2008randomized,rb2009,rb2005}. Randomized Benchmarking attempts to go beyond tomography by determining error probability per gate in computational contexts. Like GST, it is calibration free \cite{nielsen2020gate}. A further strength of randomized benchmarking is that is insensitive to variations in error between the different types of gates used \cite{helsen2019new,wallman2018randomized,proctor2017randomized,merkel2018randomized}. 

There are variations on the specific implementation \cite{rb2019different}, including numerous extensions \cite{magesan2012efficient,wallman2015estimating,kimmel2014robust,gambetta2012characterization,alexander2016randomized}. We provide a high level description of the general approach, but note that different formulations may vary on the specifics. First, a random sequence of operations (gates) is generated. Commonly, gates are chosen uniformly at random from the Clifford set, as these can be efficiently performed on a quantum processor \cite{magesan3639robust} and also can be efficiently simulated on a classical computer \cite{gottesman1997stabilizer}.  The sequence is then appended with a final operation which undoes the action of the entire sequence. For example, if the sequence has a length of $m$ and the operation (set of gates) applied at cycle $i$ is denoted as $c_i$, the effect of the first $m-1$ operations is 
\begin{equation}
C = \prod _{i=0}^{m-1} c_i
\end{equation}
\noindent where $\prod$ denotes matrix multiplication. The final operation is then chosen to be the inverse, or adjoint, of $C$ 
\begin{equation}
    c_{m} = C^{\dagger}
\end{equation}
\noindent and hence the full sequence will become the Identity operation. 
\begin{equation}
    \prod_{i=0}^m c_i= c_m C = C^{\dagger}C = \textbf{I}
\end{equation}
This final corrective operation is easy to determine, due to the Clifford set being efficient to simulate classically \cite{magesan3639robust}. Hence, a quantum state subjected to this sequence will be returned  back to its initial state (in the case of no error). This is often chosen to be the $\ket{ 00...0}$ state. In experiment, the fidelity can be found by finding the probability of measuring the initial state \cite{ibmRB}. Many sequences of the same length should be generated, and the results averaged over all cases. This process is then repeated, using sequences of various lengths. The fidelity will be a function of the sequence length and will drop exponentially as the the length increases. The fidelity can then be fit to a model, where constants will absorb the state preparation and measurement errors \cite{magesan3639robust}. This enables the extraction of the average error per gate.  For example, in a single qubit case with independent gate errors, the average fidelity of the sequence can be modeled as \cite{magesan2011scalable}

\begin{equation}
    \langle F\rangle = A(-2r+1)^m + B
\end{equation}
where $r$ is the average gate error and $A$ and $B$ are the fitting constants. This enables $r$ to depend exclusively on the fidelity of the gate operations \cite{helsen2019new}. This process can be applied to many qubit quantum states with relative ease, as the resources required scale polynomially in the number of qubits \cite{greenbaum2015introduction,magesan2011scalable,magesan2012characterizing}. As a result, it has become a standard approach to benchmarking quantum systems. The common practice of using the Clifford set, while convenient, is also a limitation, as by itself it is not universal \cite{greenbaum2015introduction}. However, strategies have been found to extend randomized benchmarking beyond the Clifford set \cite{kimmel2014robust,carignan2015characterizing,helsen2019new}. Additionally, note that the scalability of randomized benchmarking is enabled by only seeking a subset of the quantum information, it does not provide the full tomographic information about the gates \cite{merkel2013self,evans2019scalable}.  

Modern analog quantum computers have been able to simulate systems that are hard even for classical supercomputers \cite{wiebe2014hamiltonian,simon2011quantum,britton2012engineered,kim2010quantum}. This raises the concern of validating their output, as they may be untrustworthy \cite{hauke2012can,gogolin2013boson} and it is challenging to classically certify them \cite{wiebe2014hamiltonian}. \emph{Hamiltonian Learning} is a process by which the Hamiltonian of a system is estimated in order to  validate that it is simulating the correct dynamics \cite{granade2012robust,evans2019scalable,wiebe2014hamiltonian,bairey2019learning,wiebe2014quantum2}. Additionally, reconstructing the Hamiltonian of a quantum system will provide detailed diagnostic information which can enable noise reduction in experiment \cite{evans2019scalable}. The Hamiltonian of an n-qubit system, which has dimension $d=2^n$ can be described by $d^2$ parameters, though most Hamiltonians of interest can be described by $m=O(poly(n))$ \cite{evans2019scalable}. Utilizing known information about the system can significantly reduce the number of measurements required and make the process tractible \cite{granade2012robust}. \emph{Compressed quantum Hamiltonian learning} \cite{wiebe2015quantum} is a process in which the dynamics of subsystems of a large device are measured against the dynamics of a smaller system, enabling a model to be created for the larger system. Using a \emph{trusted simulator} \cite{wiebe2014hamiltonian,wiebe2014quantum2}, which has a firm known mathematical model, enables an absolute model to be generated for the larger system\cite{wiebe2015quantum}.

%It is clearly important to be able to perform quantum gates with high fidelity. It is also clearly important to be able to accurately characterize the fidelity. However, as we will see in later sections, focusing too much the fidelity of single- or two-qubit gates can be misleading. Hence, these can be surprisingly poor metrics for large-scale quantum computers.

%\redHL{Make sure that fidelity is clearly defined by the time we reach point (before first use, needless to say)}

\section{(Quantum) Computer Benchmarking}
\label{sec:computer}
This kind of benchmarking is more similar to classical benchmarking. The idea is to determine the compute capability of a quantum system running a program. This is sometimes referred to as \emph{holistic benchmarking} \cite{nielsen2020gate,nielsen2020probing}. Note that this is still considering near-term computers that do not use error correction. It shares a number of considerations with classical computing, such as latency and available parallelism. However, these metrics are not as informative for quantum computers. As to be expected, there are a number of additional considerations. 
When discussing benchmarking for full quantum computers, it is important to reconsider the role of quantum error and, if applicable, quantum error correction. 
As previously mentioned, quantum noise is not equivalent to classical noise, and these differences get more pronounced the larger the system is. 

In classical computing it would be feasible to obtain an error rate per gate and stitch these error rates together to generate an error model for a larger circuit. The quantum equivalent would be finding a fidelity for each gate, and then assuming this error rate for each gate on each qubit in the system throughout the entire program. 
%\redHL{why not differentiate btw different types of gates?} 
This is \emph{not} accurate for quantum circuits as quantum noise is context dependent \cite{WallmanISCA}. Even if good estimates of gate fidelity can be obtained, using this information to model larger systems with more qubits is not straightforward. A gate performed on one qubit may induce error in another, via quantum entanglement or physical proximity. Hence, qubits must be considered as a monolithic system and their error rates cannot be considered independently \cite{erhard2019characterizing,ferracin2019accrediting}. This is problematic as it makes it difficult to understand how noise affects large quantum computers. The whole point of creating large quantum computers is to create states that cannot be efficiently classically simulated. Unfortunately, that also means the noise becomes impossible to simulate. Hence, accurately characterizing the noise, and its significant effects on the reliability and performance, is not straightforward.

This has a few key impacts on benchmarking quantum computers. One is that it intensifies the error that is introduced when using a reduced program size as a benchmark. The error rates per qubit or per operation may be higher on a larger system. Some experimental evidence has shown that this may be overly pessimistic \cite{erhard2019characterizing}, but increased system sizes will no doubt increase susceptibility to noise due to complexity \cite{martonosi2019next}. Hence, the rate of success of a program on a small system may be significantly different from that of a larger system, and extrapolating results is not straightforward. 
%...

\subsection{Program Benchmarks}
\label{sec:programbenchmarks}
An intuitive approach is establishing a set of programs and measuring the performance of a computing system performing each one. As previously mentioned, this is common practice for classical computers. Replicating this for quantum computing would be collecting a set of quantum programs which are representative of the algorithms we would like to run on them. Common examples may be Shor's \cite{shor1999polynomial} for prime factorization or Grover's \cite{grover1996fast} for unstructured search. There are intuitive advantages to this approach, particularly in making the system perform a ``real world'' task. IonQ, a quantum start up which has an 11-qubit ion-trap quantum computer, appears to favor this approach. They tested their computer on the Bernstein-Vazirani \cite{bernstein1997quantum} and Hidden Shift \cite{van2006quantumHS,rotteler2010quantumHS} algorithms, and their metric for performance was the likelihood of measuring the correct output \cite{wright2019benchmarking}. They claim that these algorithms are representative benchmarks and the results proved their system was the best as of early 2019. 

A challenge that this approach introduces is that someone must decide which programs are important.   This introduces the issue of invested interests \cite{lilja2005measuring}. In the classical computing domain, a lot of money is on the line when benchmarking  hardware \cite{hennessy2011computer}. This problem is exacerbated by the fact that different, competing quantum technologies are superior at different programs. For example, the program benchmark approach was used in \cite{linke2017experimental}, where a handful of small quantum circuits \footnote{In literature on quantum computing, circuit refers to a sequence of quantum gates (instructions). It is analogous to ``program'' in classical computing.} were used to benchmark and compare the performance of an Ion-Trap quantum computer with a superconducting quantum computer. Amongst the chosen benchmarks were three-qubit circuits implementing the Toffoli and Margolis gates. It was found that the higher connectivity (ability of different qubits on the machine to interact) of the ion-trap computer allowed it to have a much higher relative success rate on the Toffoli circuit, which contained more two-qubit gates during the program. The success rate was more comparable on the Margolis circuit, which required fewer two-qubit gates. The authors note that how well the quantum architecture matches with the requirements of the algorithm is a major determinant of the performance \cite{linke2017experimental}. So, which of the two benchmarks are more insightful? This question highlights the difficulty of creating a representative set quantum program benchmarks. While quantum computing is more complicated, much can be learned from previously discovered pitfalls and fallacies of classical program benchmarking. Intuitively, the benchmarks will be set by the customer, rather than the manufacturer. To the extent possible, the benchmarks should reflect the applications that customers will be running on them. Another key component of accurate (and honest) benchmarking with a set of programs is transparency \cite{hennessy2011computer}. This means providing data on all the programs and not unfairly weighting some results over others. For example, a quantum benchmark suite could be created which contains many circuits which have few two-qubit interactions, but also a few prominent circuits which require many two-qubit interactions, such as the Quantum Fourier Transform (QFT) \cite{shor1999polynomial}. A quantum computer with low connectivity could be benchmarked on all circuits, with only the average performance being reported. Such analysis could falsely inflate perception of the performance.

While impressive, these modern computers are very small compared to the computers we hope to build in the coming years. Hence, these benchmarks are also very small compared to truly useful programs. While running smaller versions of real world applications introduces error, and is accepted in classical benchmarking, this is exacerbated for quantum computers. Entirely new issues may be introduced when scaling up and it is difficult to say whether measurements taken today are good indicators of future performance. For example, IonQs computer \cite{wright2019benchmarking} has all 11 qubits fully-connected, meaning each qubit can directly interact with every other qubit. This configuration is possible at this scale, but may not be for a system with hundreds or thousands of qubits. Such a system will likely require multiple fully-connected groups of qubits and communication will need to be orchestrated between them \cite{martonosi2019next}. This introduces additional complexity which is not found in these small scale benchmarks. This is analogous to the classical benchmarking of machine learning inference accelerators. An accelerator which performs well on MNIST \cite{MNISTlecun1998gradient} digit recognition will not necessarily perform well on 1000-class ImageNet \cite{imagenet_cvpr09} classification. While the problem and computation is similar, ImageNet requires much more data, and memory management becomes the bottleneck.

A more fundamental question is what quantum programs will be useful in the future. Famous algorithms such as Shor's \cite{shor1999polynomial}, Grover's \cite{grover1996fast}, and quantum chemistry \cite{ChemistryAolson2017quantum,ChemistryBmcardle2018quantum,ChemistryCcao2018quantum} are obvious examples. However, for the most part, these algorithms will remain well out of reach for some time. Currently, classical-quantum hybrid algorithms \cite{schuld2018circuit,wecker2015progress,farhi2014quantum,peruzzo2014variational} are popular due to their ability to make use of the limited resources of modern quantum computers. It is important to remember that quantum algorithm design is still an emerging field, and what actually is the best use of quantum computers is still unknown. This presents a moving target, which suggests quantum research should not too heavily invest in any one direction \cite{blume2019metrics}.
\vspace{.1cm}
\subsection{Quantifying Capability}
Considering the current limitations of quantum computers, it may be more insightful to focus on how much work a quantum computer is capable of, in contrast to performance results on specific algorithms. Because quantum technology is not mature, modern quantum computers are not yet capable of performing commercially useful algorithms. However, creating larger, more functional computers is of great interest, regardless of the applications they perform. This type of benchmarking is a shift away from traditional, classical benchmarking, and is an attempt to uniquely and objectively quantify the ``capability'' of modern quantum computers. The aim is to abstract out as much as possible, such as unique architecture characteristics and performance on specific algorithms, and create a simple metric which indicates the general computational power of the machine. While this benchmarking approach can be applied to all modern quantum computers, it is intrinsically tied to the concept of \emph{quantum supremacy} - demonstrating that a quantum computer can perform work that is not possible with classical computers. Hence, this type of benchmarking is not just used as a comparison between different points in the design space of quantum computers, but is also the  experimental assessment of the capabilities of quantum computers with respect to classical machines. Such benchmarks seek practical demonstrations of quantum computers solving well-defined problems, which are also intractable (in terms of time and hardware resources required) for classical computers. The best known classical algorithm/implementation should be used as a baseline for comparison in this case, while accounting for all sources of noise on the quantum side.

The importance of demonstrating quantum supremacy cannot be overstated. A central motivation for creating quantum computers is to solve problems which cannot be solved by other means \cite{svore2016quantum}, which will only be possible if quantum computers can achieve (and go beyond) quantum supremacy. It has been argued that this will be fundamentally impossible due to noise \cite{kalai2014gaussian}. Hence, even if an experimental demonstration of quantum supremacy does not provide scientifically or commercially useful results, it is an invaluable proof of concept for future research efforts. The success of quantum computing research is contingent upon reaching this goal. The key question then is how to quantify capability of quantum machines accurately and how to be sure when they truly achieve quantum supremacy.

Cross-Entropy benchmarking \cite{boixo2018characterizing} can be used to validate the output of a quantum computer and was created specifically as method to test for quantum supremacy. The previously described program benchmarks consider decision problems, where the measurement at the end provides an answer to a specific question. In significant contrast,  Cross-Entropy benchmarking considers \emph{sampling problems}. Here, the measurement at the end effectively allows for the sampling from a certain distribution. It is the ability of the quantum computer to create such a distribution which acts as the demonstration of quantum supremacy. Hence, it must be shown that the quantum computer produces the distribution with a sufficient fidelity and that a classical computer (using polynomial resources) cannot provide a sampling from that same distribution. Beyond demonstrating supremacy, validating sampling problems is important because they correspond to important quantum applications, such as quantum simulation.   

To produce the output distribution, a quantum circuit is constructed by choosing quantum gates (from a universal set) at random. Using a random circuit allows for limited guarantees of computational hardness, as it does not have structure \cite{arute2019quantum,bremner2016average,boixo2018characterizing,bouland2019complexity}. Executing this circuit on a quantum computer will produce a quantum state in which some states are much more likely than others, yet is also widely distributed over the $2^n$ possible measurement outcomes, where $n$ is the number of qubits. The distribution resembles a speckled intensity pattern produced by light interference in laser scatter \cite{arute2019quantum}. For a large quantum computer, $2^n$ will be much greater than the number of samples (executions of the circuit) that can be taken. Hence, it is very unlikely that two different measurements will be same \cite{boixo2018characterizing}. This makes it difficult to discern it from a uniform random number generator by simply looking at the samples. However, the output can be distinguished and validated if the circuit is also classically simulated \cite{lloyd2013pure,popescu2006entanglement,gogolin2013boson,ududec2013information}. The classical simulation (which requires exponential resources on a classical computer) will provide the exact output quantum state, and hence also the exact probabilities of measuring each result (in the absence of noise). This allows the comparison of the output of the experimental quantum circuit with the ideal quantum circuit, and also with any classical algorithm that attempts to generate the same distribution. The information theory definition of entropy is \cite{shannon2001mathematical}  
\begin{equation}
    H(X) = - \sum_{i=0}^{N-1}\; P(x_i)log P(x_i)
\end{equation}

where $X$ is a random variable which can take on $N$ values and $P(x_i)$ is the probability of observing $X = x_i$. The output of the quantum circuit, and a classical algorithm attempting to emulate it, is a random variable which can take $2^n = N$ values. The cross-entropy between the distribution of the ideal quantum probabilities ($p_U$) and the polynomial classical algorithm probabilities ($p_{pcl}$) can be defined as \cite{boixo2018characterizing} 

\begin{equation}
    H(p_{pcl},p_U) = - \sum _{j=1}^N \;p_{pcl}(x_j|U) \;log\; p_U(x_j)
\end{equation}

where $p_U(x_j)$ is the probability that the quantum circuit, $U$, will produce the output $x_j$ and $p_{pcl}(x_j|U)$ is the probability that the classical algorithm emulating $U$ will also produce $x_j$. The value of interest is the average quality of the classical algorithm, which can be found by averaging the cross-entropy over an ensemble of random quantum circuits, $\set U$:
\begin{equation}
    E_U[H(p_{pcl},p_U)] = E_U \left[ \sum_{j=1}^N p_{pcl}(x_j|U)\frac{1}{log \;p_U(x_j)}\right ]
\end{equation}
%It was argued in \cite{boixo2018characterizing} that the polynomial classical algorithm cannot accurately reproduce the distribution. A single error in the quantum circuit (such as an extra X or Z gate error) will produce an output distribution which is nearly entirely uncorrelated with $p_U$. Hence, any method which uses approximation (which introduces error) will also produce a distribution that is nearly statistically uncorrelated with $p_U$. The cross-entropy of the ideal quantum circuit output with the output of the polynomial classical algorithm will be the same as the cross-entropy with the output of a uniform random number generator, denoted as $H_0$ 
\b{It was argued in \cite{boixo2018characterizing,bouland2019complexity,bremner2016average} that the polynomial classical algorithm cannot accurately reproduce the distribution. }

\b{When performing the circuit on a quantum computer, a single error (such as X or Z error) will cause the output to become nearly statistically uncorrelated with $p_U$. Hence, depolarizing noise will cause an increase in the entropy of the output of the quantum computer, and it will resemble that of a uniform distribution. This means that the quantum computer, which is subject to noise, will only barely be able to produce a superior cross-entropy to that of a uniform random number generator. }
\begin{equation}
    E_U[ H(\b{Uniform},p_U)] = log \;N + \gamma = H_0
\end{equation}
where $\gamma = 0.577$ is the Euler constant. On this concept the quantum supremacy test is based. Any classical or quantum algorithm, $A$, which produces bitstring $x_j$ with probability $p_A(x_j|U)$, can be evaluated on how well it predicts the output of an ideal quantum random circuit, $U$, by comparing its cross-entropy relative to that of a uniform classical sampler. This metric is called the cross-entropy difference \cite{boixo2018characterizing}: 

\begin{equation}
    \Delta H(p_A) = H_0 - H(p_A,p_U) = \sum_j \left ( \frac 1 N - p_A (x_j|U)  \right ) \; log \; \frac{1}{p_U(x_j)}
\end{equation}

If the algorithm A produces the distribution perfectly, with no errors, the cross-entropy difference will be 1. If the algorithm A produces an uncorrelated distribution, the cross-entropy difference will be 0 \cite{arute2019quantum,boixo2018characterizing}. Effectively, this is a measure of how consistent the outcomes are with the predicted probabilities \cite{blume2019volumetric}. A classical algorithm (using polynomial resources) should fail, and a quantum algorithm running on a sufficiently powerful quantum computer should succeed. The experimentally determined cross-entropy difference, denoted by $\alpha$, can be found by taking $m$ samples from the quantum computer, with each producing a bit string $x^{exp}$, and then using classical simulation to find the value $p_U(x^{exp})$. $\alpha$ is then estimated by \cite{boixo2018characterizing} 

\begin{equation}
    \alpha ~= H_0 - \frac 1 m \sum_{j=1}^m log \frac{1}{p_U(x_j^{exp})}
\end{equation}

Quantum supremacy is achieved if a quantum computer can produce a higher $\alpha$ than a classical computer can. Unfortunately, quantum supremacy also implies that the $p_U$ values required to determine $\alpha$, which are produced by classical simulation (using exponential resources), can no longer be computed by a classical computer in a reasonable amount of time. Hence, measuring $\alpha$ directly is not possible if the quantum computer truly has achieved quantum supremacy. However, it should be possible to extrapolate $\alpha$, if it has been reliably found at sizes which are slightly below the limit of quantum supremacy ($p_U$ can still be computed by sufficiently powerful classical supercomputers \cite{boixo2018characterizing}). \b{In the summer of 2019 Google used cross-entropy benchmarking \footnote{Google used linear cross-entropy benchmarking, which differs from the standard cross-entropy benchmarking process explained here. However, the argument is similar for why linear cross-entropy benchmarking is easy for a quantum computer and hard for a classical computer.} to quantify the capability of their 52-qubit quantum computer}, Sycamore, where they claim to have demonstrated quantum supremacy \cite{arute2019quantum}. It is also possible to use cross-entropy benchmarking to find the average fidelity of individual gates \cite{arute2019quantum,boixo2018characterizing}, similar to randomized benchmarking.

 Note that tests for quantum supremacy doesn't necessarily require a universal set of gates on the quantum computer. For example, in the case of Boson Sampling, recent work has demonstrated that these algorithms can be composed by notably more noise-tolerant (yet not necessarily universal) set of gates \cite{bosonSampling}. 

A similar and influential sampling-based benchmark is quantum volume from IBM \cite{bishop2017quantum,quantumvolume2}. The process of determining quantum volume shares many of the steps with cross-entropy benchmarking. According to the authors \cite{bishop2017quantum}, there are 4 factors which determine the capability of a quantum machine:
\begin{enumerate}
    \item The number of physical qubits
    \item The number of gates that can be applied before errors make the the output unreliable
    \item The connectivity of the machine
    \item The number of operations that can be run in parallel
\end{enumerate}
It is assumed that quantum volume targets modern, noisy quantum computers. Hence, factor 2) is referring to running a quantum algorithm without error correction and it is assumed there is an upper limit on the number of gates possible. In the future, it will be of more interest to determine whether the error rate is low enough to enable efficient error correction, or how often error correction needs to be applied. Hence, quantum volume will need to be adapted or superseded in the future \cite{quantumvolume2}.

Similar to cross-entropy benchmarking, quantum volume attempts to abstract out all considerations and generates a single number which quantifies the capability of a quantum computer. The idea is to measure something which can be improved by each of the 4 considerations, meaning systems that are superior in each consideration will generally achieve a higher quantum volume. The score is determined by the largest random quantum circuit a quantum computer is able to complete successfully. Whereas cross-entropy benchmarking uses the cross-entropy to quantify the validity of the output, quantum volume uses the heavy output generation problem \cite{aaronson2016complexity}. Heavy output generation also requires classical exponential time to validate the output from the quantum computer, as it relies on the ability to classically simulate the quantum operations \cite{aaronson2016complexity}. Again, this will not be possible once physical quantum computers reach the scale where their quantum states are not efficiently classically simulable. By simulating the quantum operation, the output measurement probabilities can be exactly determined. When ordering all possible outputs from most likely to least likely, all outputs which have a greater probability than the median are called \emph{heavy outputs}. The quantum circuit is then repeatedly performed and the output measured. If the measurements produce \emph{heavy outputs} more than 2/3 of the time, the quantum computer ``passes''; otherwise it fails. If noise has completely destroyed the information in the quantum state, heavy outputs will be generated 50\% of the time \cite{quantumvolume2}. The largest quantum circuit that the computer can get a pass on corresponds to the quantum volume, $V_Q$ \cite{quantumvolume2} 
\begin{equation}
    log_2 V_Q = argmax_m \left (\;\; min(m,d(m)) \;\;\right )
\end{equation}

 where $m$ is the number of qubits and $d(m)$ is the maximum achievable depth of an $m$-qubit circuit. Hence, quantum volume is the largest square ($m=d$) circuit that can be successfully run.

Quantum volume has some limitations. A number of which were addressed in \cite{blume2019volumetric}, which proposes a framework called Volumetric Benchmarks. The idea is to make quantum volume more general, by allowing different shapes of circuits, kinds of circuits (random,  periodic, subroutines of algorithms), and different criteria for success. As noted by the authors of \cite{blume2019volumetric}, errors will affect different kinds of circuits differently, such as coherent errors getting magnified by periodic circuits but getting smeared out by random circuits. This provides evidence that universal benchmarks should be avoided. While quantum volume is a novel and useful concept, it uses exclusively random circuits. As we show in our simulations, random circuits have a similar affect to applying randomized compiling \cite{wallman2016noise}. Another potential drawback is that quantum volume can produce potentially misleading results. For example, a Honeywell quantum computer achieves a high quantum volume score due to its extremely high fidelity operations \cite{stutz2020trapped}. However, given the relatively low qubit count, its dynamics can be simulated classically with ease \cite{aaronsonblog}.

Another significant approach is that of Cycle Benchmarking \cite{erhard2019characterizing}. This is similar to the process of randomized benchmarking. The core idea is to break a quantum program into sets of operations on all the qubits (cycles) and then individually characterize the fidelity of each cycle. This allows one to quantify how well the computer can do specific operations. Additionally, arbitrary programs can be broken into a finite set of cycles \cite{WallmanISCA}. This prevents the number of required characterizations from growing exponentially with the number of qubits. Using cycle benchmarking, the ``benchmarks'' would be individual cycles and the metric is the process fidelity, as defined in Equation \ref{eq:fidelity}.

%\vspace{.1cm}
%\subsection{\b{A Note on Quantum Supremacy}}
%\input{supreme.tex}

\section{Fault Tolerant (Quantum) Computer Benchmarking}
\label{sec:faulttolerant}
Thus far we've considered qubit benchmarking and holistic computer benchmarking for near-term quantum machines. While there has been much effort to create practically useful quantum algorithms for these machines, as of yet their performance falls far short of their classical competitors. Hence, the main goal of these near-term demonstrations has been for research and  to learn how to achieve scalability.  Once computers are built that have sufficient qubit counts and sufficiently low error rates, full-scale quantum error correction (QEC) \cite{greenbaum2015introduction} will become possible. QEC can detect and correct errors during the run of a program, and hence enable computers to perform arbitrarily long quantum programs. This is referred to as fault-tolerant quantum computing \cite{preskill1998fault}. Such computers will be capable of solving real-world problems with much greater speed than classical computers, performing well beyond the threshold for quantum supremacy. Naturally, such fault-tolerant computers will significantly change the landscape of benchmarking as it breaks many of the assumptions of previously described techniques. For example, the maximum circuit depth is effectively infinite, hence a metric such as quantum volume would be limited only by qubit count. More fundamentally, the quantum states of these machines would be much too large to simulate, and hence the classical validation of the heavy output generation problem would not be possible. A typical use case for large-scale fault-tolerant quantum computers will be running algorithms which will be impossible to run on a classical computer, but for which the output can efficiently be checked for correctness on one \cite{loceff2015course}, such as Shor's algorithm \cite{shor1999polynomial}.  

However, no fault-tolerant computers have been built to date, and hence there is not much research dedicated to benchmarking them. Likely, quantum benchmarking will look much more similar to classical benchmarking, where the emphasis will transition to performance (latency to produce the final result) amongst quantum processors which have sufficiently many qubits to perform the algorithm of interest. Benchmarking may take a similar form as program benchmarks, as described in Section \ref{sec:programbenchmarks}, except the chosen benchmarks would be real-world, full-size applications rather than sample toy cases. Additionally, unlike the modern computers performing the program benchmarks described in Section \ref{sec:programbenchmarks}, a fault-tolerant computer should not fail due to noise. Hence, the probability of success will not be a representative metric.

The performance of fault-tolerant computers will have a strong dependence on the method QEC used as it has a high overhead. QEC utilizes many qubits \cite{fowler2012surface}, involves many, repeated quantum operations, and also requires significant classical hardware resources to manage its orchestration \cite{metodi2006quantum}. Both the qubit chip \cite{duckering2020virtualized} and the supporting classical architecture \cite{tannu2017taming} will need to be specifically designed to support QEC. For an introduction to quantum error correction, we refer the interested reader to \cite{gottesman2010introduction}.

%\vspace{.4cm}
\section{Simulations}
%\redHL{if space permits, add a benchmark table w/ brief description, acronyms, may be inputs, may be gate related stuff etc. }}
\label{sec:experiments}
To illustrate the impact of the different types of quantum noise in different computational contexts, and the resulting difficulty of choosing representative benchmarks, we run representative key quantum algorithms at sizes that are experimentally feasible. We use a Quantum Adder \cite{cheng2002quantumadder}, the Quantum Fourier Transform (QFT) \cite{shor1999polynomial}, and the Quantum Approximation Optimization Algorithm (QAOA) \cite{farhi2014quantum}. In addition,  we use an idle circuit (which has no computational gates) and a random circuit (composed of random X, Y, Z, H, and CNOT gates), 
%\redHL{make sure all covered in primer}, 
for reference. Note that this is not the same randomized circuit as used in Quantum Volume \cite{quantumvolume2}, which generates a set of arbitrary random unitary operations, which need to be decomposed into a universal gate set. For our random circuit, we want to view the effects of performing our gate set in random fashion, without introducing the complexity of gate compilation (which is needed for our other algorithms).  Unless otherwise stated, our metric is the process fidelity \cite{flammia2011direct} described in Equation \ref{eq:fidelity}. We note that other metrics may be equally suitable. However, we chose process fidelity as it is widely used in the quantum information science community \cite{gilchrist2005distance} and it shines light on the complexities of benchmarking mentioned earlier. 

The QFT and QAOA benchmarks contain gates which are precise rotations around the X- or Z-axis of the bloch sphere. $R_z(\theta)$ is a rotation around the Z-axis by angle $\theta$ and $R_x(\theta)$ is the equivalent for the X-axis. Currently available modern quantum computers, such as IBM's machines \cite{Qiskit}, can perform these operations directly. This can be powerful, as it enables quantum computation to occur relatively quickly, which 
%will 
also mitigates 
%against 
the effects of noise. As this gate set is likely to be used for some time into the future, we include it in our simulations of the QFT and QAOA. However, this gate set is not compatible with randomized compiling (RC) or quantum error correction. Hence, it is not scalable to the level of fault-tolerant quantum computers. For progressing into the fault-tolerant regime, the Clifford+T set is a good option. This gate set is universal and enables use of RC. Hence, we include Clifford+T versions of the QFT and QAOA as well. While we are using the Clifford+T gate set, we are not performing error correction in the simulation. Hence, these do not represent fault-tolerant computations. However, using this gate set can provide insight on how the impact of noise is affected by the chosen gate set. Each gate in Clifford+T can be implemented with the previously mentioned X- and Z- rotations -- using the Clifford-T set just restricts the angles used. When performing these gates on physical qubits, as we do here, the operations are not changed in any fundamental way. While we do not show simulations of error correction, it should be noted that, in order to perform error correction, groups of physical qubits would form a single logical qubit, and the logical operations consist of many individual gates on the physical qubits \cite{fowler2012surface}. 

The drawback of Clifford+T is that precise rotations must be broken into sequences of gates which can perform the operation approximately. This increases the length of the circuit. For single qubit gates, we use the gridsynth decomposition method from \cite{selinger2012efficient} which finds an approximation to Z rotations with Hadamard (H), T and S gates. This is sufficient, as any single qubit gate $U$ can be implemented using the Euler angles \cite{selinger2012efficient}
\begin{equation}
U = R_z(\beta)\;R_x(\gamma)\;R_z(\delta) =R_z(\beta)\;H\;R_z(\gamma)\;H\;R_z(\delta)
\end{equation}
for some $\beta$, $\gamma$, and $\delta$. Hence, any quantum gate can be approximated by a sequence of approximate $R_z$ gates and H gates. An additional complication exists in that controlled versions of these gates cannot be implemented directly \cite{kliuchnikov2012fast}. For a general case, controlled 2-qubit versions of the gates can be implemented by including an additional ancilla (scratch) qubit and using an alternative sequence of gates \cite{amy2013meet}. However, both the QFT and QAOA can be implemented with just CNOT gates and single qubit $R_z$ gates. Hence, controlled- H, S, and T gates are not required.

The Quantum Adder performs binary addition using a sequence of quantum full-adders on two input integers which are basis state encoded (1 qubit for each bit). If performing n-bit addition, the quantum adder requires $3n+1$ qubits. We perform 2-bit addition (using 7 qubits). The adder only uses gates in the Clifford+T set and hence does not require gate decomposition. {It does use the Toffoli (doubly controlled X gate), which we implement with a sequence of CNOT, H, and T gates.}

The QFT effectively performs the Discrete-time Fourier Transform (DFT) on the amplitudes of the input quantum state, and is the core of Shor's algorithm. The width and depth of the circuit are determined by the number of input qubits. We use 4 input qubits. The circuit contains Hadamard gates and controlled-Z rotations, which can be implemented with CNOT and $R_z$ gates. 

QAOA has gained a lot of attention recently as it is considered to be a strong candidate to demonstrate quantum supremacy. \b{QAOA is a \emph{variational} algorithm, where quantum computation is used in tandem with classical optimization \cite{wecker2015progress}. }It is commonly applied to the Max-Cut problem. \b{The input to Max-Cut is a graph which represents a combinatorial optimization problem. Vertices are variables and the edges between vertices are constraints. The goal is to partition the vertices into two sets, maximizing the number of edges between the sets. The \emph{maximum cut} is the number of edges between the sets in a best possible partition, which is the most constraints that can be satisfied simultaneously.} Max-Cut is NP-Hard, however, it is \emph{not} expected that quantum computers will be able to solve NP-Hard problems in polynomial time. QAOA only provides an approximate solution and currently has worse performance than classical approximate algorithms \cite{goemans1995improved,kugel2010improved}. \b{QAOA consists of a sequence of stages. Each stage consists of controlled-Z rotations (determined by the input graph) followed by single qubit X rotations.} The hyperparameter $p$ determines the number of stages in the quantum circuit. For each stage, there is a parameter $\gamma$, which determines the angle of the controlled-Z rotation, and a parameter $\beta$, which determines the angle of the X rotations. The main challenge of QAOA is to determine an optimal set of parameters in order to produce good output \cite{guerreschi2019qaoa}. When implemented on a real quantum computer, optimization algorithms such as SPSA \cite{spall1992multivariate} or gradient descent can be used to update the parameters \cite{sung2020using}. This requires many optimization passes, where each pass requires many (thousands or millions of) samples of the quantum circuit output. The number of samples required heavily depends on the problem size and noise level in the system. Here, we consider only a pre-optimized circuit with $p=1$. When $p=1$, the optimal circuit is guaranteed to produce an output with an expectation value that is at least 0.6924 times the \b{maximum cut} \cite{farhi2014quantum,grove} on 3-regular graphs. Here, we use a 3-regular input graph with 8 vertices (represented by 8 qubits), which has a maximum cut of 12. This means QAOA should produce an expectation value of 8.3 in the case of no error. For this problem, we use expectation value of the output as the metric. As QAOA is designed to also work on a small scale, it can be used effectively with continuously parametrized rotation gates available on near term machines. We implement QAOA with these gates as well as with the Clifford+T set. A useful tutorial for QAOA is provided in \cite{xanaduqaoa}.

For our simulations, we assume the quantum computer has an all-to-all connectivity and full parallelism. This means 2-qubit gates can be performed without any overhead for movement, and multiple single qubit gates can be performed simultaneously, given that they do not operate on the same qubits. Moving qubits with swap gates in order to compute on machines with limited connectivity is a well-studied problem \cite{li2019tackling,zulehner2018efficient}. Our observations here will also apply, but there will be further complicating factors depending on the specific topology. For discussion on how noise and topology interact see \cite{tannu2019not}.  
%\redHL{some insight here?} 

We also view the impact of RC \cite{wallman2016noise} in each of our simulations. RC was designed as a method to convert coherent noise effectively into stochastic Pauli noise. This is significant, as coherent noise can be much more destructive and is predominantly seen in physical experiments. Hence, RC not only provides a significant noise mitigation technique, but also maintains the validity of previous theorems and results which have been generated by assuming stochastic Pauli noise. We emphasize, as noted by the original authors \cite{wallman2016noise}, that RC is not designed to have any impact on a Pauli noise model, hence we expect to see no improvement in the fidelity if a Pauli noise model is used. Additionally, for most of our simulations we are using process fidelity as the figure of merit, for which RC is not expected to add as much benefit. RC will provide greater impact if using a norm-based measure, such as the trace distance in Equation \ref{eq:tracedistance}. In order to perform RC, we need to divide the gate set into ``easy'' and ``hard'' gates. {Easy and hard can generally be thought of as the difficulty in implementing the gate, such as the expected error rate, but the essential requirement is that the physical noise on the easy gates be independent of which easy gate is performed. } We follow \cite{wallman2016noise} and set easy gates as the Pauli gates and the phase gate S; where the hard 
%\st{single qubit} 
gates are H, T, and all 2-qubit gates.  
%\redHL{insight on what easy and hard entail?} 
However, instead of controlled-Z, we use CNOT. Nearly all gates in the decomposed circuits are ``hard'' gates. Hence, it is necessary to interleave these gates with idle cycles in order to implement RC. From a high level, simulation perspective, this initially seems to imply that the cost of RC is a near doubling of the circuit length. However, as noted by the authors of \cite{wallman2016noise}, on real hardware (such as ion traps and superconductors) entangling operations, which are required for ``hard'' multiple qubit gates, must be inserted between local gates. These local gates, which are required even without RC, can be ``compiled into'' the randomized gates. Hence, in practice RC can be implemented with no additional circuit length overhead. Therefore, to make our simulations more representative of physical experiments, we insert the additional idle operations into our circuits whether RC is performed or not.

\begin{table}[]
%\scalebox{.95}{
    \centering
    \begin{tabular}{|c|c|c|c|c|c|}
        \hline
         Benchmark & \# Qubits & Logical & Physical & Gate & Metric\\
          &  &  Depth & Depth &  Set &\\
         \hline
         IDLE & 4 & 2-70 & 2-70 & I & Process Fidelity \\
         \hline
         RANDOM & 4 & 2-70 & 2-70 & I,X,Y,Z,H,CNOT & Process Fidelity\\
         \hline
         Adder & 7 & 30 & 30 & H, T, CNOT & Process Fidelity\\
         \hline
         QFT & 4 & 10 & 10 & H,$R_z$,CNOT & Process Fidelity \\
         \hline
         QFT (Clifford+T)  & 4 & 10 & 229 &  H, T, S, CNOT& Process Fidelity \\
         %\hline
         %Quantum & 3 & 7 & 525 (0), & X, H, T, S, CNOT & Accuracy \\
         %Classifier &  &  & 537 (1), & &  \\
         %(0,1,2)& & & 430 (2) & & \\
         \hline
         QAOA & 8 & 10 & 10 & H, $R_z$ ,$R_x$, CNOT & Expectation Value\\
         \hline
         QAOA (Clifford+T) & 8 & 10 & 113 & H, T, S, CNOT & Expectation Value\\
         \hline
    \end{tabular}
%    }
     \vspace{.1cm}
    \caption{Benchmarks used in simulations. \# Qubits is the number of qubits required for each input size. Logical depth is the depth of the quantum circuit (number of sequential gates) required before compilation into Clifford+T set. Physical depth is the depth after compilation. The Adder and QFT have two different input sizes. }
    \label{tab:my_label}
\end{table}

{ We use four representative noise models which fall into different categories discussed in Section \ref{sec:noise}.}  While no quantum system will have a single source of noise, we use them in isolation to demonstrate how the nature of the noise (along with the assumptions made in noise models) significantly impacts the results.
%\redHL{1. terminology: error model vs. noise model; 2. should be tied to previous discussion on noise} 
The first is a standard Pauli noise, where the probability of X, Y, and Z errors are all equally likely. While the most commonly used noise model, it generally provides the worst (and overly optimistic) estimates or error correcting threshold error rates \cite{gutierrez2015comparison}. Pauli noise is a non-unitary and incoherent model. To model purely coherent and unitary noise, we follow the same approach as in  \cite{bravyi2018correctingcoherenterrors}, where we assume constant Z-rotations by angle $\theta$ for each qubit, $(e^{i\theta Z})^{\otimes n}$, for various values of $\theta$. While this model makes some simplifying assumptions, it is representative. The third is a combination of Pauli and coherent noise, where we follow the model in \cite{greenbaum2017modelingcoherent}. This includes static X rotations and X Pauli errors. The fourth error model is Amplitude Damping, which is a commonly used and realistic noise model. It models loss of energy from the system to the environment and is a non-unitary process. {We sweep the noise over a range of values which are similar to experimental error rates and are expected in near term computers, as listed in Table \ref{tab:errorrates}. Each noise type is injected in every qubit in every cycle, regardless if the qubit is operated on or not. We assume the same noise rates for single and two-qubit gates. While two-qubit gates will typically have a higher rate of noise in a physical experiment, our values are swept over ranges typically seen for both single- and two-qubit gates.} Unless otherwise stated, our metric of choice is the process fidelity of the noisy operation, $\tilde G$, to the noiseless operation, $G$, \cite{erhard2019characterizing,flammia2011direct}. If the noisy operation $\tilde G$ is free of error, the process fidelity will be 1. We repeat experiments with different input pure states. We generate the input states at random in the same manner as \cite{barnes2017quantum}, by selecting random polar coordinates on the Bloch sphere for each qubit and report the average process fidelity.

Numerous quantum simulators exist,
\cite{haner2016high,steiger2018projectq,svore2018q,javadiabhari2014scaffcc,roetteler2017design},
many of which would be suitable to run our simulations. However, as we are implementing algorithms and incorporating noise models from a variety of sources, and did not want to unintentionally bias our experiments by relying on any specific software, we chose to run our simulations with the statistical programming language R \cite{R}. This allows us to fully and independently define our experiments. Additionally, R is highly optimized to perform matrix multiplication, which is the essential component for density matrix simulation. Our source code is available at \cite{code}, which is an extension of our R package QuantumOps \cite{quantumops}.
%Note that the binary code from \cite{gridsynth} is required 
To perform the gate decomposition, we rely on the gridsynth algorithm \cite{gridsynth}.

Performing simulation with density matrices allows us to avoid much Monte Carlo simulation. However, Monte Carlo simulation is still needed due to the randomness introduced by our random input states, use of RC, and the random circuits for the random circuit benchmark.

\begin{table}[]
%\scalebox{.95}{
    \centering
    \begin{tabular}{|c|c|c|c|c|}
    \hline
         Error & Pauli & Coherent & Pauli+Coherent & Amplitude \\
          Level &  &  &  &  Damping \\
         \hline
         0 & 0 & 0$\pi$ & 0 , 0$\pi$ & 0 \\
         \hline
         1 & 0.01 & $\pi/30$ & 0.01 , $\pi/30$ & 0.01\\
         \hline
         2 & 0.02 & $\pi/15$ & 0.02 , $\pi/15$ & 0.02 \\
         \hline
         3 & 0.03 & $\pi/10$ & 0.03 , $\pi/10$ & 0.03 \\
         \hline
    \end{tabular}
%    }
    \vspace{.1cm}
    \caption{Noise levels tested for each noise type. Noise is inserted in every qubit in every cycle, regardless if it is being operated on. Pauli noise rate refers to probability of inserting and X, Y, or Z gate. Coherent noise is the rotation angle applied. For Pauli+Coherent, only X Pauli gates and X rotations are applied to align with the model in \cite{greenbaum2017modelingcoherent}. Amplitude Damping error rate refers to the parameter $\gamma$ \cite{tomita2014low}, which is the probability of relaxation to the ground state.}
    \label{tab:errorrates}
%    \vspace{.7cm}
\end{table}

\section{Results}
\label{sec:results}

\begin{figure*}[]
\centering
%\subfloat[Pauli Noise]{\includegraphics[width= 2.5in]{figures/varyIdle_Pauli_4qubits.jpg}}
\subfloat[Pauli Noise]{\includegraphics[width=.25\textwidth]{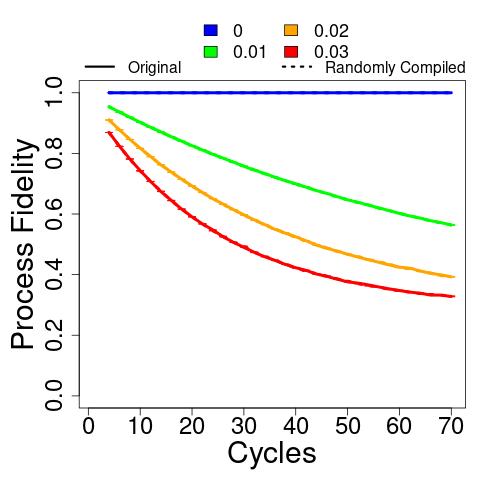}}
\subfloat[Coherent Noise]{\includegraphics[width=.25\textwidth]{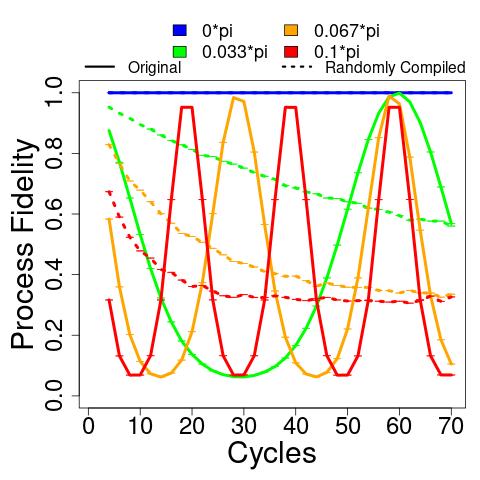}}
%\\
\subfloat[Pauli+Coherent Noise]{\includegraphics[width=.25\textwidth]{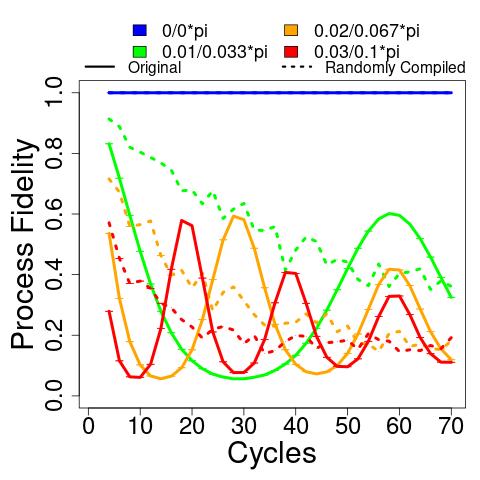}}
\subfloat[Amplitude Damping]{\includegraphics[width=.25\textwidth]{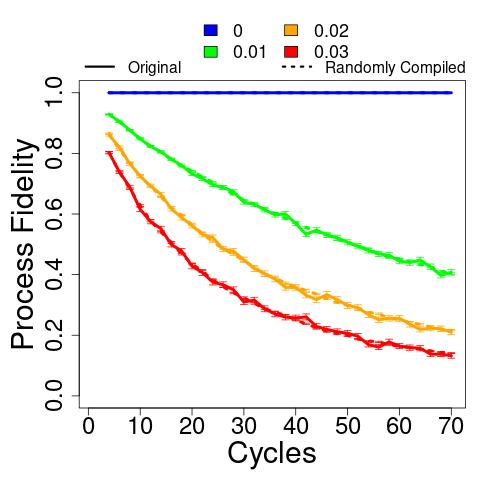}}
%\\
\caption{Idle circuit with 4 qubits under various noise models.}
\label{fig:idle}
\end{figure*}

\begin{figure*}[]
\centering
\subfloat[Pauli Noise]{\includegraphics[width=.25\textwidth]{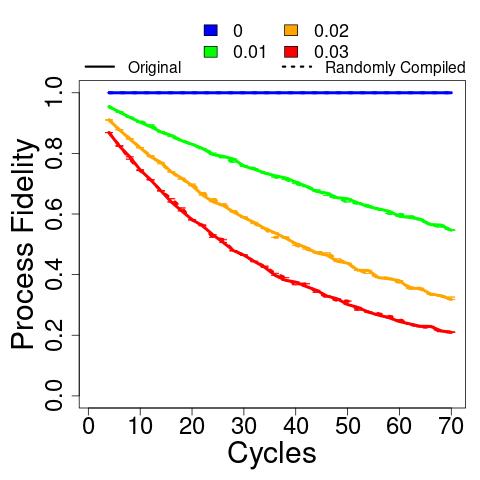}}
\subfloat[Coherent Noise]{\includegraphics[width=.25\textwidth]{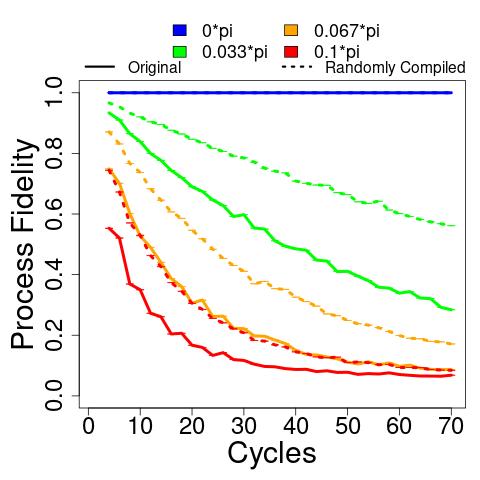}}
%\\
\subfloat[Pauli+Coherent Noise]{\includegraphics[width=.25\textwidth]{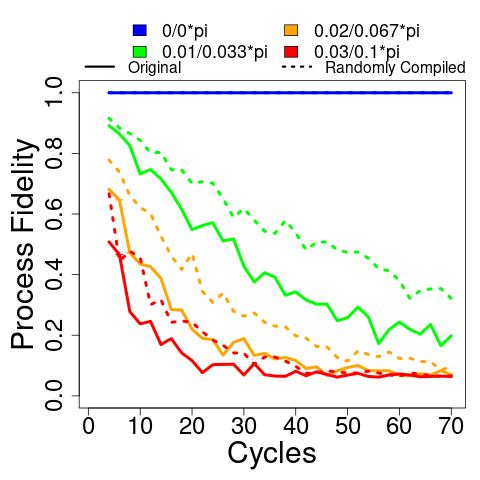}}
\subfloat[Amplitude Damping]{\includegraphics[width=.25\textwidth]{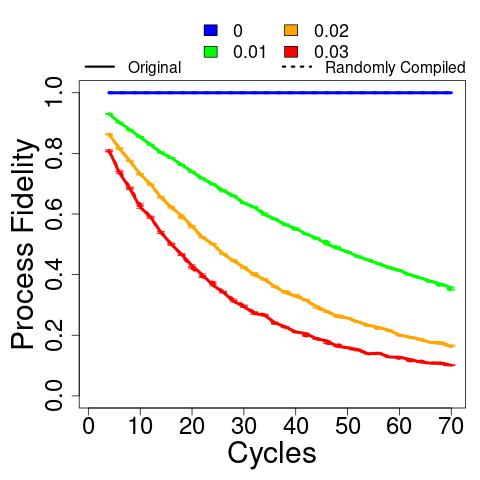}}
%\\
\caption{Random circuit with 4 qubits under various noise models.}
\label{fig:random}
%\vspace{.5cm}
\end{figure*}

\begin{figure*}[]
\centering
\subfloat[Pauli Noise]{\includegraphics[width=.25\textwidth]{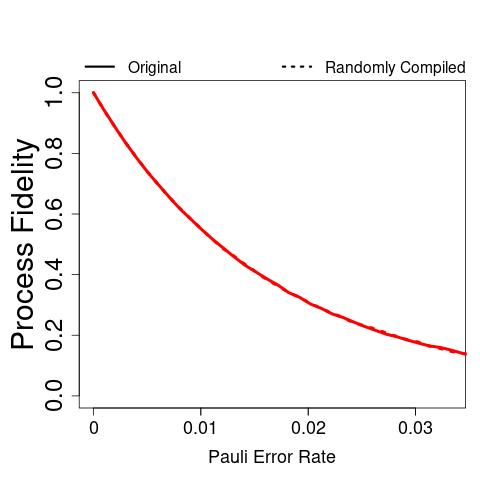}}
\subfloat[Coherent Noise]{\includegraphics[width=.25\textwidth]{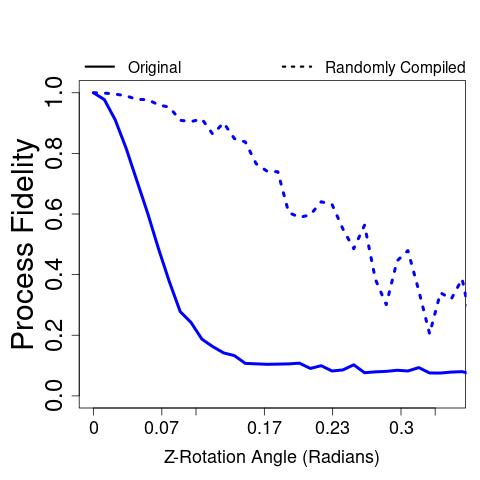}}
%\\
\subfloat[Pauli+Coherent Noise]{\includegraphics[width=.25\textwidth]{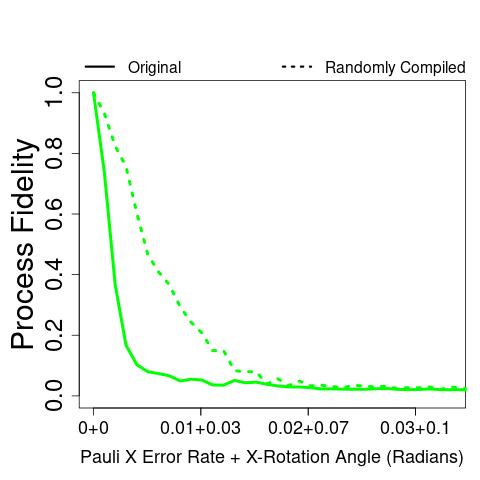}}
\subfloat[Amplitude Damping]{\includegraphics[width=.25\textwidth]{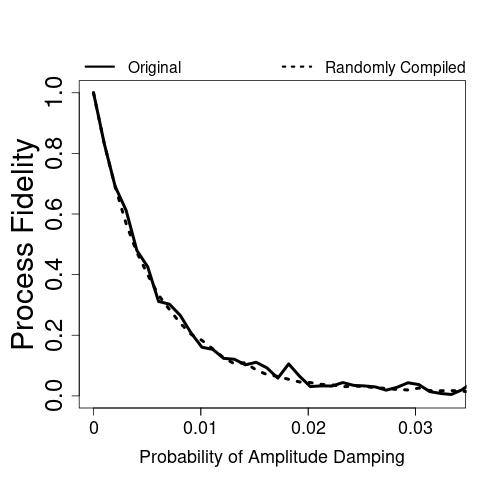}}
%\\
\caption{2-bit addition circuit under various noise models.}
\label{fig:addition}
%\vspace{.4cm}
\end{figure*}

\begin{figure*}[]
\centering
\subfloat[Pauli Noise]{\includegraphics[width=.25\textwidth]{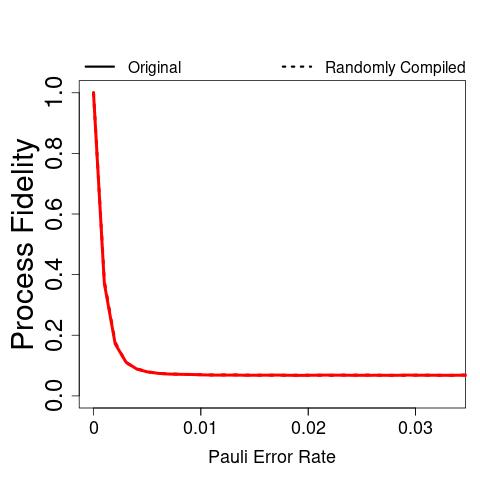}}
\subfloat[Coherent Noise]{\includegraphics[width=.25\textwidth]{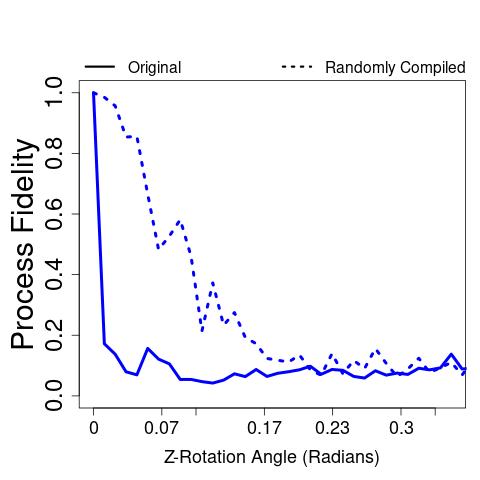}}
%\\
\subfloat[Pauli+Coherent Noise]{\includegraphics[width=.25\textwidth]{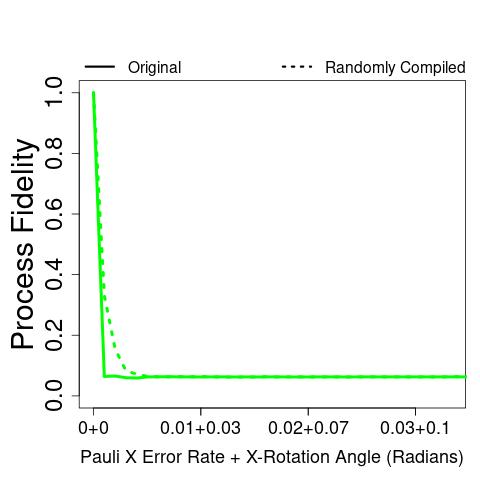}}
\subfloat[Amplitude Damping]{\includegraphics[width=.25\textwidth]{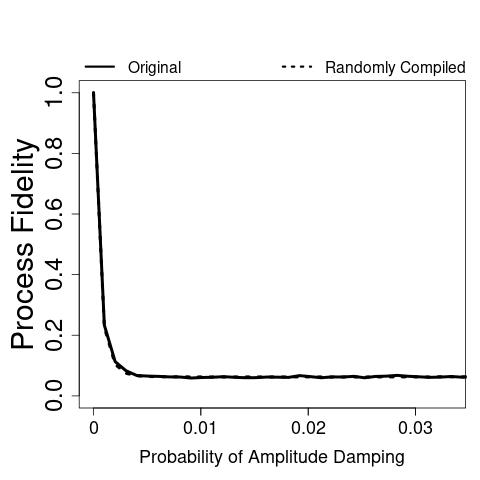}}
%\\
\caption{4-qubit QFT (Clifford+T) under various noise models.}
\label{fig:qft}
%\vspace{.5cm}
\end{figure*}

\begin{figure*}[]
\centering
\subfloat[Pauli Noise]{\includegraphics[width=.25\textwidth]{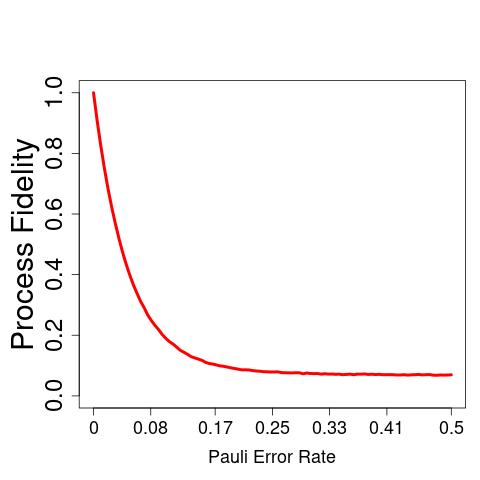}}
\subfloat[Coherent Noise]{\includegraphics[width=.25\textwidth]{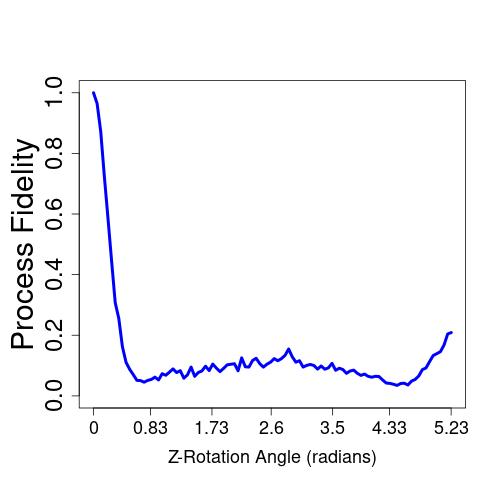}}
%\\
\subfloat[Pauli+Coherent Noise]{\includegraphics[width=.25\textwidth]{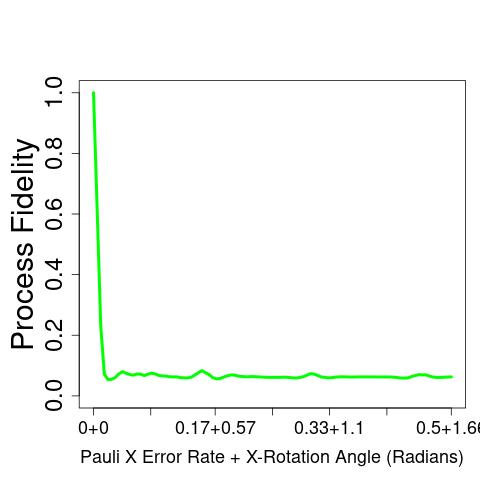}}
\subfloat[Amplitude Damping]{\includegraphics[width=.25\textwidth]{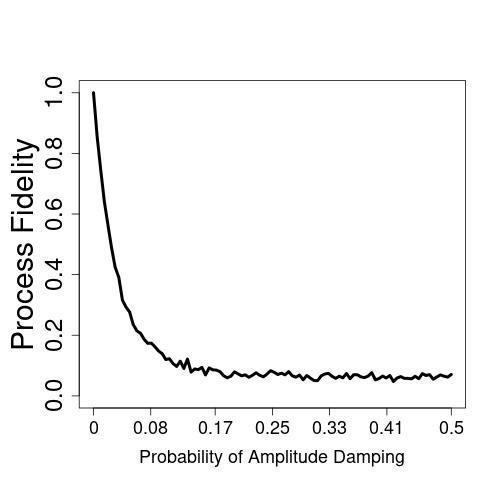}}
%\\
\caption{4-qubit QFT with parameterized rotation gates under various noise models.}
\label{fig:decomposedQFT}
% \vspace{.4cm}
\end{figure*}

\begin{figure*}[]
\centering
\subfloat[Pauli Noise]{\includegraphics[width=.25\textwidth]{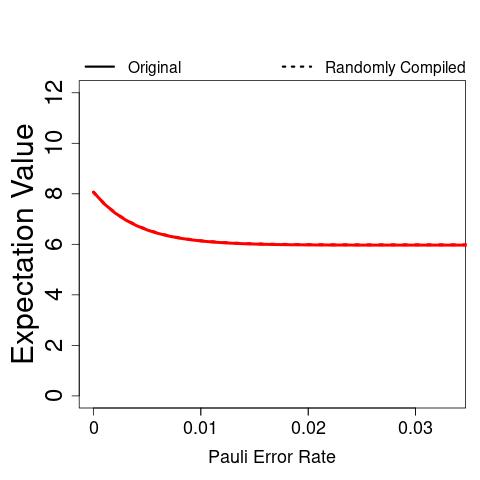}}
\subfloat[Coherent Noise]{\includegraphics[width=.25\textwidth]{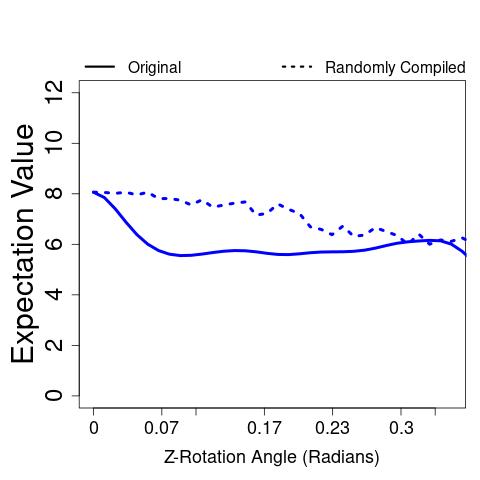}}
%\\
\subfloat[Pauli+Coherent Noise]{\includegraphics[width=.25\textwidth]{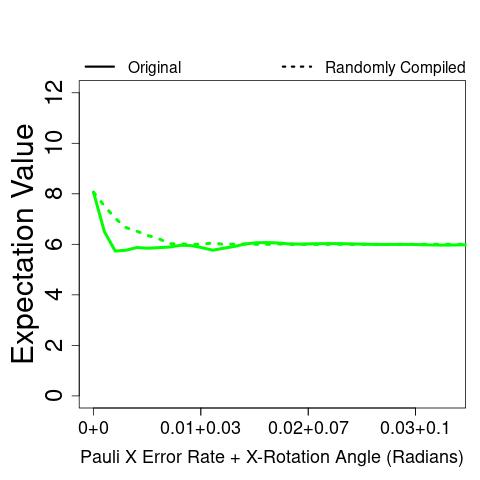}}
\subfloat[Amplitude Damping]{\includegraphics[width=.25\textwidth]{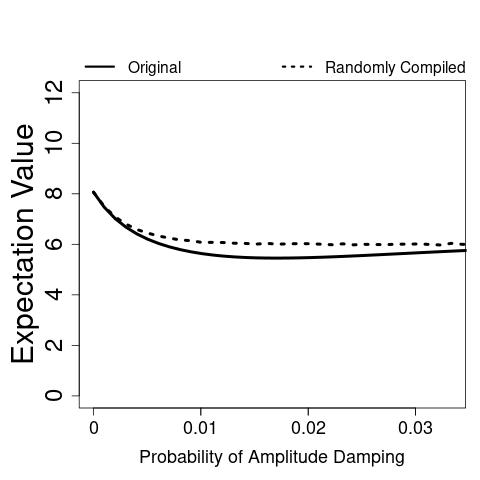}}
%\\
\caption{QAOA (Clifford+T) expectation value under different noise models.}
\label{fig:qaoa}
%\vspace{.4cm}
\end{figure*}

\begin{figure*}[]
\centering
\subfloat[Pauli Noise]{\includegraphics[width=.25\textwidth]{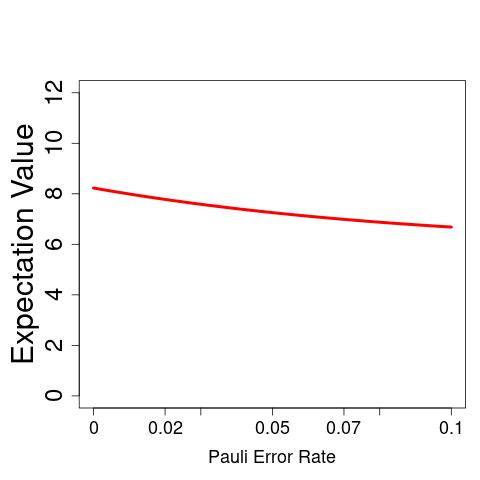}}
\subfloat[Coherent Noise]{\includegraphics[width=.25\textwidth]{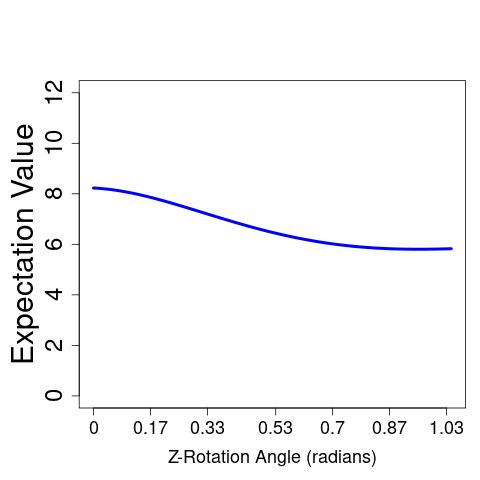}}
%\\
\subfloat[Pauli+Coherent Noise]{\includegraphics[width=.25\textwidth]{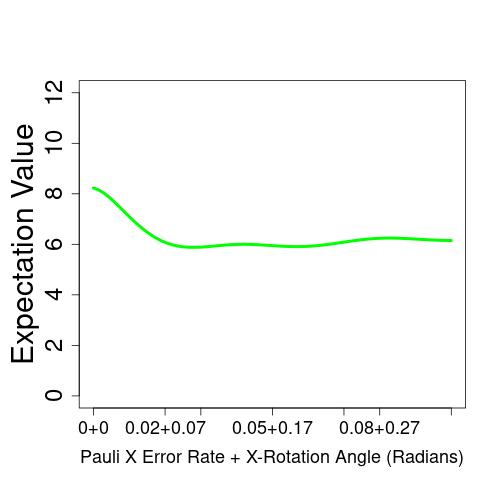}}
\subfloat[Amplitude Damping]{\includegraphics[width=.25\textwidth]{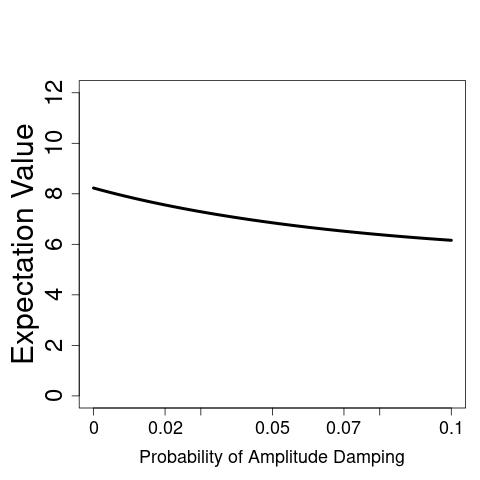}}
%\\
\caption{QAOA with parameterized rotation gates expectation value under different noise models. As the circuit is significantly shorter than Clifford+T QAOA, the noise is tested over a significantly larger range.}
\label{fig:qaoaRotation}
%\vspace{.4cm}
\end{figure*}

For the Idle and  Random circuits we sweep the {noise levels over the values listed in Table \ref{tab:errorrates}} and plot the effect on the fidelity for circuits of different lengths. As the Addition, QFT, and QAOA circuits have a constant depth, we sweep the 4 error models over a fine-grained range and plot the process fidelity or accuracy of the output versus the error rate. 
 
Results for the Idle circuit are shown in Figure \ref{fig:idle} and results for the Random circuit are shown in Figure \ref{fig:random}. Both circuits are performed from 2 to 70 cycles. Cycles here indicate the length of a single gate. Note that error level 0 shows a process fidelity of 1, meaning there is no corruption of the quantum state. It is highly noticeable how coherent noise affects the Idle and Random circuits differently. For the Idle circuit, coherent noise has an immediate drastic impact on the fidelity. We must note the strange behavior of the Idle circuit under coherent noise. Due to our simplified coherent noise model, the fidelity returns with a periodicity determined by the constant angle of rotation. This is unlikely to be a physically realistic phenomenon, and even if it was, \emph{it would not be possible to exploit this fact unless one was completely aware of the exact effects of the physical noise}. As physical noise is difficult to model and predict, it is highly unlikely one would have such knowledge. Note that randomized compiling removes this periodic effect.

The true observation from the simulation of coherent noise on the Idle circuit is the immediate destructive nature of even a slight coherent noise source. Note that randomized compiling mitigates this impact and causes the coherent noise to have the same effect as stochastic Pauli noise. Interestingly, this same coherent noise is not as destructive on the Random circuit. In fact, even without Randomized Compiling, the error decays exponentially just as it does under stochastic Pauli noise. This finding is consistent with \cite{blume2019volumetric}, which says that coherent noise will affect randomized circuits much differently than idle or cyclic circuits. The coherent noise gets ``smeared out'' by the randomization inherent in a random circuit. Additionally, this suggests that the randomized benchmark circuits in Quantum Volume \cite{quantumvolume2}, depending on the gate decomposition, may not be representative of many quantum circuits. Noteworthy is that Randomized Compiling may not be necessary if the circuit already contains a high degree of randomness. However, the vast majority of useful quantum algorithms do not have such structure. %The results for the Quantum Classifier add further credence to this concept. When classifying the IRIS data set, the classifier remains highly immune to coherent errors. This is in part due to a natural robustness due to the simplified output, but also due to the short length of the circuit. The longer Adult classifying circuit is affected more, but not as significantly as the Idle circuit.

\vspace{.1cm}
\noindent\boxed{
\begin{minipage}[t]{0.99\columnwidth}
The chosen quantum noise model has a drastic impact on the performance of quantum algorithms. Hence, one must be sure that the assumptions on the noise present in a physical system are appropriate. Additionally, the effect of the quantum noise is largely determined by the nature of the quantum algorithm being performed. Hence, one must be cautious when choosing quantum algorithms for benchmarks.
\end{minipage}
}
\vspace{.1cm}

Results for the addition circuit are shown in Figure \ref{fig:addition}, the Clifford+T version of QFT in Figure \ref{fig:decomposedQFT}, and the parameterized rotation gate version of QFT in Figure \ref{fig:qft}. Note that circuits using the parameterized rotation gates cannot be randomly compiled. Randomized Compiling produces a significant increase in fidelity when coherent noise is present. 

Note that 
%when the noise is 
under Pauli noise or Amplitude Damping, Randomized Compiling does not provide a significant improvement. For Pauli noise, as it is already entirely random, it remains unmodified by the random compilation. This is noted by the inventors of RC \cite{wallman2016noise}. As RC is designed to mitigate coherent rotations of the qubit state, it is also intuitive that it would not significantly mitigate Amplitude Damping, which models energy loss of the system. However, RC may help in some specific circumstances, such as when a qubit is held in the excited $\ket{1}$ for an extended period of time. As Amplitude Damping models the collapse from $\ket 1$ to $\ket 0$, the $\ket 1$ state is more vulnerable. As noted by the authors of \cite{saki2019study}, Amplitude Damping is more destructive if the quantum data in a program contains more qubits in the $\ket 1$ state. RC could make the state oscillate, reducing the amount of time the qubit will spend in the excited, more vulnerable state. However, this condition was not present in our benchmarks.

%In addition to the noise model, the effect of Randomized Compiling varies depending on the algorithm. While it does not improve fidelity under Pauli noise for any algorithm, it does significantly improve fidelity of the the Idle circuit and QFT when coherent noise is present. However, it is again unhelpful or harmful for the Quantum Classifier. 

\vspace{.1cm}
\noindent \boxed{
\begin{minipage}[t]{.99\columnwidth}
While a critical tool to enable scalable quantum computing, especially for modern machines, Randomized Compiling (RC) cannot be used as a generalized noise mitigation technique. As noted by the original paper \cite{wallman2016noise}, randomized compiling is used to ``convert'' coherent noise to stochastic Pauli noise. This is significant, as modern, physical quantum computers are dominated by coherent errors. Additionally, this conversion to Pauli noise maintains the validity of previously established proofs which have assumed Pauli noise. \end{minipage}
}
\vspace{.1cm}

%Noticeably, the Quantum-Classifier suffers significantly from all forms of noise. Before performing gate decomposition, the classifier proved to be quite robust, as intuition suggested. Unfortunately, the precise rotations required by the classifier resulted in long sequences of gates for approximation, which significantly increases vulnerability to noise. The gridsynth \cite{gridsynth} algorithm allows for a wide range of precision in the approximation. We achieved the best results when significantly reducing the approximation precision in order to achieve a shorter circuit. Due to this approximation, the classifier suffers an accuracy loss even when no noise is present. This demonstrates a trade-off between precision and robustness to noise. Consistent with previous experiments, Randomized Compiling helps only when coherent noise is present. At sufficiently high levels of  noise, the classification of the Quantum Classifier can become effectively a random guess. The accuracy of IRIS converges to 1/3 as there are 3 equally likely classes. In the case of Amplitude Damping, where at higher levels the state is guaranteed to be in the $\ket 0$ state, the accuracy decays to 2/3. This is because, as the task is to identify 1 of the 3 possible classes, always guessing ``no'' is correct 2/3 of the time.

Results for the Clifford+T QAOA are shown in Figure \ref{fig:qaoa}. If no error is present, the QAOA circuit will produce an expectation value of 8.3, as this is 0.6924$\times$ of the maximum cut (12), which is expected for a QAOA circuit with $p=1$. The expectation value of a random guess for the max-cut problem is at least 50\% that of the maximum cut \cite{grove}. For the specific problem we used, a random guess produced an expectation value exactly 50\% of that maximum. Hence, increasing the noise, which increases the entropy of the output, generally caused the expectation value to decay to 6. Noticeably, the expectation value drops to that of a random guess even for very low levels of noise. Amplitude Damping noise has significant potential to drop the expectation value below 50\%, as high levels of this noise will cause the state to transition more towards the $\ket{00...0}$ state. The $\ket{00...0}$ state does not satisfy any of the input problem's constraints and hence produces an expectation value of 0. Notice that RC prevents this drop below 50\%.

QAOA using parametrized rotation gates is shown in Figure \ref{fig:qaoaRotation}. Due to not using decomposition, the circuit is much shorter and hence significantly less impacted by noise. Note that this gate set is not compatible with randomized compiling or quantum error correction. Hence, it is not scalable but still feasible for the small cases which we are considering. Despite the difference in gate sets, this version of QAOA shows similar expectation value patterns due to the different noise types. The expectation value tends to converge towards a value that is 50\% of the maximum. Again, a combination of Pauli and coherent noise is particularly destructive.  
\section{Conclusion}
\label{sec:conc}
%\redHL{1. Summary what this paper is; 2. Summarize take-aways}
Computer architecture uses layers of abstraction to manage complex problems. This approach has already been applied to quantum computing. Unfortunately, quantum systems are notorious at defying abstraction and simplifying assumptions. It is easy to make invalid assumptions and generate inaccurate results. Here, we showed that quantum noise is more complex and difficult to model than is often assumed. This has profound effects everywhere, and can be felt significantly even at higher levels of the system stack. This complicates the task of benchmarking, which is already challenging and full of subtlety for classical computers. The noise model, the target application, and the performance metric all need to be carefully considered. 

%%%%%%% -- PAPER CONTENT ENDS -- %%%%%%%%

%%%%%%%%% -- BIB STYLE AND FILE -- %%%%%%%%
\bibliographystyle{unsrt}
\bibliography{ref}
%%%%%%%%%%%%%%%%%%%%%%%%%%%%%%%%%%%%

\end{document}